\documentclass[a4paper]{aa}

\usepackage[varg]{txfonts}
\usepackage[english]{babel}
\usepackage{graphicx}	
\usepackage{amsmath}	
\usepackage{amssymb}

\usepackage{epsfig}
\usepackage{url}
\usepackage{multicol}
\usepackage{multirow} 

\usepackage{natbib}
\bibpunct{(}{)}{;}{a}{}{,} 
\usepackage{hyperref}

\begin{document}

\title{Influence of galaxy stellar mass and observed wavelength on disc breaks in S$^4$G, NIRS0S, and SDSS data
    \thanks{Tables 1 and 3 are only available in electronic form at the CDS via anonymous ftp to cdsarc.u-strasbg.fr (130.79.128.5) or via http://cdsweb.u-strasbg.fr/cgi-bin/qcat?J/A+A/}}

\author{Jarkko Laine\inst{1}
\and Eija Laurikainen\inst{1}
\and Heikki Salo\inst{1}}

\institute{Astronomy Research Unit, FIN-90014 University of Oulu, P.O. Box 3000, Oulu, Finland \email{jarkko.laine@oulu.fi} }

\date{Received XXXX / Accepted YYYY}

\titlerunning{Disc breaks across masses and wavelengths}
\authorrunning{Laine et al.}

\abstract
{Breaks in the surface brightness profiles in the outer regions of galactic discs are thought to have formed by various internal (e.g. bar resonances) and external (e.g. galaxy merging) processes. By studying the disc breaks we aim to better understand what processes are responsible for the evolution of the outer discs of galaxies, and galaxies in general.}
{We use a large well-defined sample to study how common the disc breaks are, and whether their properties depend on galaxy mass. By using both optical and infrared data we study whether the observed wavelength affects the break features as a function of galaxy mass and Hubble type.}
{We studied the properties of galaxy discs using radial surface brightness profiles of 753 galaxies, obtained from the $3.6 \, \mu m$ images of the Spitzer Survey of Stellar Structure in Galaxies (S$^4$G), and the $K_{\rm s}$-band data from the Near InfraRed S0-Sa galaxy Survey (NIRS0S), covering a wide range of galaxy morphologies  ($-2 \le T \le 9$) and stellar masses ($8.5 \lesssim \log_{10} (M_{*}/M_{\sun}) \lesssim 11$). In addition, optical Sloan Digital Sky Survey (SDSS) or Liverpool telescope data was used for 480 of these galaxies.}
{We find that in low-mass galaxies the single exponential profiles (Type I) are most common, and that their fraction decreases with increasing galaxy stellar mass. The fraction of down-bending (Type II) profiles increases with stellar mass, possibly due to more common occurrence of bar resonance structures. The up-bending (Type III) profiles are also more common in massive galaxies. The observed wavelength affects the scalelength of the disc of every profile type. Especially the scalelength of the inner disc ($h_{\rm i}$) of Type II profiles increases from infrared to \emph{u}-band on average by a factor of $\sim 2.2$. Consistent with the previous studies, but with a higher statistical significance, we find that Type II outer disc scalelengths ($h_{\rm o}$) in late-type and low mass  galaxies ($T > 4$, $\log_{10} (M_{*}/M_{\sun}) \lesssim 10.5$) are shorter in bluer wavelengths, possibly due to stellar radial migration populating the outer discs with older stars. In Type III profiles $h_{\rm o}$ are larger in the \emph{u} band, hinting to the presence of young stellar population in the outer disc. While the observed wavelength affects the disc parameters, it does not significantly affect the profile type classification in our sample. Our results indicate that the observed wavelength is a significant factor when determining the profile types in very low mass dwarf galaxies, for which more Type II profiles have been previously found using optical data. }{}


\maketitle


\section{Introduction}

Multiple structures observed in galaxies provide us with information of how galaxies have evolved to their current state. Galactic discs are generally thought to be the first structures formed in the collapse of gas and subsequent star formation in the dark matter halo (e.g. \citealt{white1978,springel2005}), leading to exponentially decreasing surface density profiles (e.g. \citealt{dalcanton1997,mo1998}). Since then galaxies have undergone complex internal and external processes creating new, and modifying existing structures that we observe in low redshift galaxies. The discs show a multitude of signatures of these processes, such as bars, spirals, rings, lenses, and tidal features. Many of these structures are manifested as deviations from simple exponential behaviour of the surface brightnesses of the discs, known as disc breaks\footnote{Disc \emph{truncations} are thought to be separate feature observed in edge-on galaxies and further out in the discs compared to breaks, see \cite{martinnavarro2012}.}. In fact, it has been established by \cite{erwin2005}, \cite{pohlen2006}, \cite{gutierrez2011}, and \cite{laine2014}, that single exponential discs (generally designated as Type I) form only a minority among the bright galaxies. These studies have also shown that discs with two exponential sections, with a down-bending break (Type II) between the sections, appear roughly in half of the bright galaxies, and up-bending breaks (Type III) in one third of them. Naturally, the exact division between the types depends on how small deviation in the slope is taken to indicate a break.

Discovery of these different disc profile types have raised the question of how much of the disc properties are original, and how much they are modified in later evolutionary processes. Some of the formation theories of Type II profiles do indeed favour the idea that the discs are formed with a break, for example due to angular momentum conservation of the collapsing gas in the initial galaxy formation \citep{vanderkruit1987}. However, a majority of the theories assume that the discs are initially formed exponential, and the breaks result from later evolutionary processes. Most conclusive formation theories have been developed for Type II breaks, for which the favoured formation mechanism also depends on the morphological type of the galaxy. In early-type disc galaxies (i.e. $T \lesssim 2$) the breaks are commonly associated with outer rings and lenses (Type II-OLR, \citealt{pohlen2006,erwin2008,gutierrez2011}), with the radius of the structure matching well with the break radius \citep{laine2014}. These breaks are thought to be associated to the outer Lindblad resonances of bars, where the outer rings and lenses are formed in galaxies. In late-type galaxies  (i.e. $T \gtrsim 2$) majority of the Type II breaks are assumed to have formed when a star formation threshold is reached, beyond which radius star formation is reduced (Type II-CT, e.g. \citealt{schaye2004,elmegreen2006,roskar2008a,sanchezblazquez2009,martinezserrano2009}). Furthermore, the outer disc beyond the break can be populated by stellar radial migration, which could be induced by transient spiral structure (e.g. \citealt{sellwood2002,roskar2008a,martinezserrano2009,roskar2012}), by bar-spiral resonance coupling (e.g. \citealt{minchev2010}), or by satellite galaxy interactions (e.g. \citealt{quillen2009,bird2012}). In the simulations of \cite{debattista2006}, the angular momentum redistribution and radial migration caused by the bar-spiral resonant coupling was linked to the actual formation of a break in the outer disc, with observational support for this process found by \cite{munozmateos2013}. Radial migration is a slow random-walk process and therefore the properties of the break should depend also on the observed stellar populations: the effects are expected to be more pronounced in the older stellar populations. Alternative formation scenarios for Type II breaks have also been suggested. Namely, the down-bending breaks that are inside or at the bar radius (Type II.i) might be formed by redistribution of matter caused by the bar \citep{pohlen2006,erwin2008,kim2016}. Furthermore, some of the Type II breaks at the disc outskirts could arise from global asymmetries of the outer discs affecting the azimuthally averaged surface brightness profiles (Type II-AB, \citealt{pohlen2006,erwin2008}).

Type III break-formation models favour environmental origins. In simulations, both minor \citep{laurikainen2001,penarrubia2006,younger2007} and major merger \citep{borlaff2014} events have been shown to spread the outer disc producing the up-bending profile. Alternatively, gas accretion could perturb the outer disc forming a Type III profile \citep{minchev2012}. Indeed, recent observational studies have provided support for the external origin of Type III profiles. For example, \citet{head2015} found, using two-dimensional decompositions of Coma cluster galaxies, that galaxies with Type III profiles show an increase in the bulge size and luminosity, compatible with a merger scenario where bulge growth is accompanied with a Type III break formation. \citet{elichemoral2015} studied various scaling relations of Type III discs, but did not find any significant differences between galaxy types or between barred and unbarred galaxies. Their results support a dynamical response in galaxies, possibly due to galaxy interactions, as a formation mechanism. Some indication of environmental origin has also been found in studies where correlations between the parameters describing the properties of the local environment of Type III galaxies, and the properties of the profiles are found \citep{pohlen2006,laine2014}. However, Type III profiles are not always explained by environmental effects. Specifically, \citet{maltby2012b} suggested that even 15 \% of the up-bending profiles could be explained by an excess of bulge light at large radii. Also, at lower surface brightness levels the halo light \citep{bakos2012,martinnavarro2014,watkins2016} or thick disc \citep{comeron2012} could contribute to the surface brightness profile, in a manner that causes an apparent up-bend in the outer disc.

In particular, Type II breaks have been studied in stellar age and colour profiles (e.g. \emph{g'-r'} in \citealt{bakos2008}). The average Type II colour and age profiles show inside-out blueing or younger stellar ages, with the bluest colours and youngest stars being found at the break radius, beyond which the colour again reddens as the stellar populations become older. This U-shaped profile has been explained by the star-formation threshold creating the break, while the outer disc reddening would be due to radially migrated old stars (e.g. \citealt{roskar2008a}). Such a clear connection with the disc evolution and break formation has not been found for Type I or III profiles. Several authors have studied the average behaviour of colour and age profiles with varying samples and techniques, and have reached similar conclusions (\citealt{bakos2008,azzollini2008a,yoachim2010,yoachim2012,roediger2012,zheng2015,marino2015} and references therein). Exceptions have also been found so that some Type II discs  have flat colour and age profiles, in a similar manner as those in Type I discs \citep{yoachim2012,roediger2012,ruizlara2015}. And vice verse, some of the Type I discs have a U-shaped profile usually associated with Type II discs \citep{ruizlara2015}. However, despite the behaviour of the colour profiles obtained in the literature, how the basic parameters of the discs change as a function of the observed wavelength has not yet been studied. In addition, possible morphological-type dependency of the colour profiles has not been examined, even though the Type II break-formation mechanism has been suggested to be different in early- and late-type galaxies.

So far, not much attention has been paid to how galaxy mass affects the different processes responsible for the disc profile types. Galaxies in the opposite ends of the mass scale seem to have formed the bulk of their mass at different epochs in the Universe, which should also affect the formation of their discs: while high mass galaxies grow inside-out as more material is accreted to the disc, low mass galaxies shrink as the feedback processes are more efficient in getting rid of the gas (e.g. \citealt{perez2013,pan2015}). \cite{herrmann2013} found that in many parameters the breaks form a continuation from the low mass dwarfs to high mass disc galaxies. For example, the break radius  systematically increases with galaxy stellar mass or with the absolute magnitude of the galaxy (e.g. \citealt{pohlen2006,munozmateos2013,herrmann2013}). \cite{herrmann2013} also found that the down-bending Type II breaks are significantly more common among the dwarf galaxies. However, a majority of their sample consisted of dwarf irregular galaxies, leaving a gap from the high mass disc galaxies to the normal low mass disc galaxies, which have not yet been studied.

In this paper we expand the study of \cite{laine2014}, and study the properties of the disc breaks in radial surface brightness profiles of a sample of 753 galaxies, using $3.6 \, \mu m$  data from the Spitzer Survey of Stellar Structure in Galaxies (S$^4$G, \citealt{sheth2010}), and $K_{\rm s}$-band data from the Near Infrared S0-Sa galaxy Survey (NIRS0S, \citealt{laurikainen2011}). This combined sample covers a wider range of morphological types and stellar masses than studied before, and bridges previous studies that have separately focused on high and low mass galaxies.  We aim to explore what effects the stellar mass has on the basic properties of the disc breaks, for example, how common they are, and how the basic parameters change as a function of stellar mass. In addition, we use optical SDSS/Liverpool Telescope data from \citet{knapen2014} for 480 of these galaxies to study how the disc parameters change as a function of wavelength band.


\begin{figure*}
  \begin{center}
    \includegraphics[width=\textwidth]{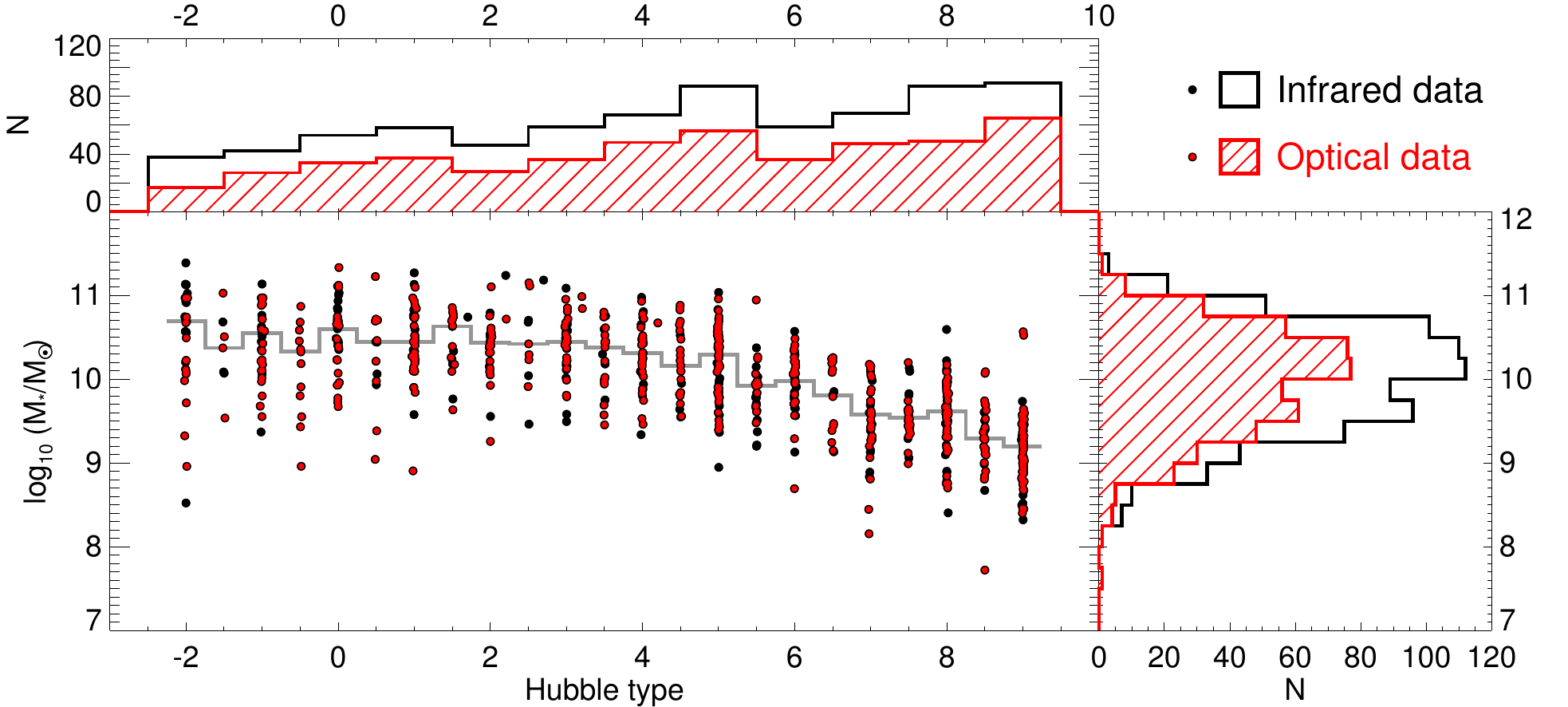}
    \caption{Stellar mass of the sample galaxies as a function of the Hubble type. In addition, histograms of the Hubble type (top) and stellar mass (right) distributions are shown. Galaxies with infrared data and galaxies with optical data in at least one waveband are shown separately in all panels. The grey line in the main panel shows the median mass in each Hubble type bin, and defines the division to `high' and `low' mass galaxies used in section \ref{results:wave-mass}.}
    \label{sample_histo}
  \end{center}
\end{figure*}

\begin{figure}[!h]
  \begin{center}
    \resizebox{\hsize}{!}{\includegraphics{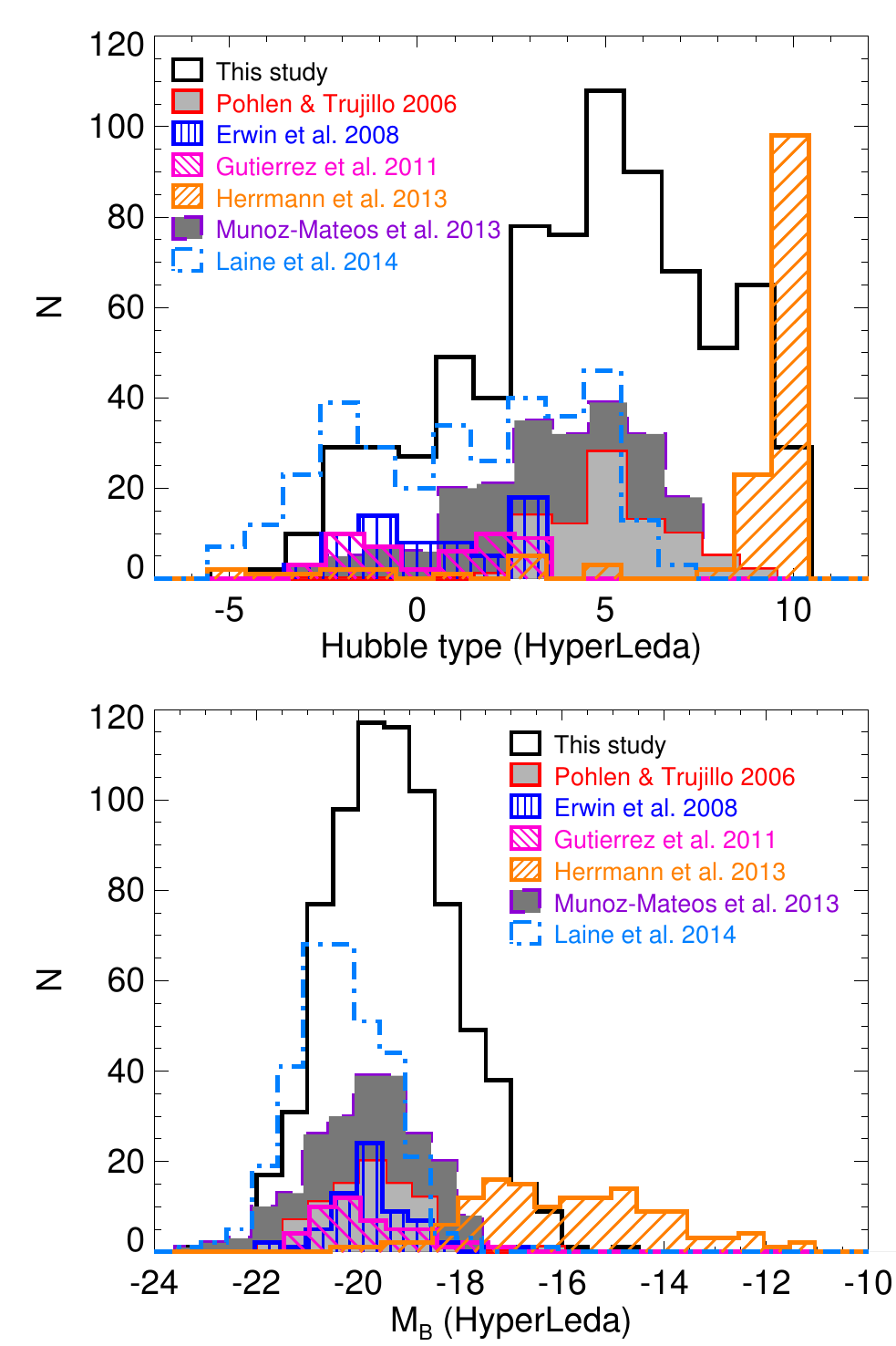}}
    \caption{A comparison of our sample with some previous disc profile studies \citep{pohlen2006,erwin2008,gutierrez2011,herrmann2013,munozmateos2013,laine2014}. In the \emph{upper panel} we show the Hubble types and in the \emph{lower panel} the absolute B-band magnitudes. Note that the values are taken from HyperLeda \citep{makarov2014}.}
    \label{other_studies}
  \end{center}
\end{figure}

\section{Sample and the data}
\label{sample-selection}

\subsection{Sample}

We used two infrared surveys as a base for our study, which together provide a good coverage of the whole Hubble sequence and cover a wide range in stellar masses of galaxies ($-5 \lesssim T \lesssim 11$, $7.5 \lesssim log_{10} ( M_* / M_{\sun}) \lesssim 11.5$). These surveys are the Spitzer Survey of Stellar Structure in Galaxies (S$^4$G, \citealt{sheth2010}) and the Near InfraRed S0-Sa galaxy Survey (NIRS0S, \citealt{laurikainen2011}). From these surveys we selected galaxies that fulfil the following criteria:
\begin{center}
  \begin{itemize}
   \item Hubble type $-2 \le T \le 9$, (using the mid-IR classification in \citealt{buta2015}, and \citealt{laurikainen2011}, for S$^4$G and NIRS0S, respectively),
   \item minor/major axis ratio of the outer disc $b/a > 0.5$.
  \end{itemize} 
\end{center}
The disc axis ratio criterion is applied to restrict to nearly face-on or moderately inclined galaxies, for which the surface brightness profile shapes can be determined in a reliable manner. The Hubble type restriction is applied to exclude Hubble types for which our methods are unreliable. This sample also extends the study of \citet{laine2014} to lower galaxy stellar masses and later Hubble types. In their study the lower Hubble type limit is set to $T \geq -3 $, but in this study the bulge-dominated $T = -3$ types are excluded because it is difficult to separate the bulge and disc contributions in the radial surface brightness profiles. In the later Hubble types our current study extends the study of \cite{laine2014} by including types $T=8$ and $T=9$. The latest types ($T>9$) usually have a patchy irregular morphology and do not show a well-defined centre, thus complicating the disc profile studies. The initial sample selection criteria returns 952 galaxies. After inspecting the data we additionally dismissed 199 galaxies, for which the study of the disc profile shape is unreliable (e.g. bright saturated stars near the discs, image artefacts, insufficient image depth for the study of the outer disc). The final sample consisted of 753 galaxies with infrared data (28 from NIRS0S, 725 from S$^4$G). Of these galaxies 262 were included in the sample of \citet{laine2014}, but for consistency were reanalyzed here. In Figure \ref{sample_histo} we show the masses of the sample galaxies as a function of Hubble type. We also show the median mass in each Hubble type bin, which is used to divide galaxies to `high' and `low' masses in section \ref{results:wave-mass}. In Figure \ref{other_studies} we compare our sample to some previous disc break studies.

\subsection{Data}

For the S$^4$G galaxies we used the 3.6 $\mu m$ images, and the $K_{\rm s}$-band images for the NIRS0S galaxies. 
For the S$^4$G images we used the foreground star and artefact masks and galaxy stellar masses from \citet{munozmateos2015}. The parameters for the galaxy centres, orientations, ellipticities, background sky levels, and standard deviations of the background sky $(\sigma_{\rm sky}$) are from \citet{salo2015}. The mid-IR morphological classifications of the galaxies in the S$^4$G are from \citet{buta2015}. For the NIRS0S galaxies we used the cleaned and flux calibrated $K_{\rm s}$-band images from \citet{laurikainen2011}, where the values of $\sigma_{\rm sky}$, galaxy centres, position angles, disc inclinations, and the morphological classifications are also given.

Optical data is added using the compilation of \citet{knapen2014}, who collected and prepared the Sloan Digital Sky Survey (SDSS) images in the \emph{u}, \emph{g}, \emph{r}, \emph{i}, and \emph{z} bands from Data Releases 7 and 8 (DR7 \citealt{abazajian2009}; DR8 \citealt{aihara2011}), for 1768 S$^4$G galaxies. \cite{knapen2014} have also acquired data with the Liverpool telescope in the \emph{g}-band for 111 S$^4$G galaxies which do not have SDSS data. Optical data in all five SDSS bands is available for 450 galaxies in our infrared sample. For 30 galaxies Liverpool telescope \emph{g}-band data is available, resulting in 480 galaxies having \emph{g}-band data. Three of the galaxies with $K_{\rm s}$-band data from NIRS0S have data in the five SDSS bands. Properties of the sample galaxies, and available bands are included in Table \ref{galaxy_properties}, which is available in full form at the CDS.

\begin{figure*}
    \centering  
    \begin{minipage}{0.46\textwidth}
     \resizebox{\hsize}{!}{\includegraphics{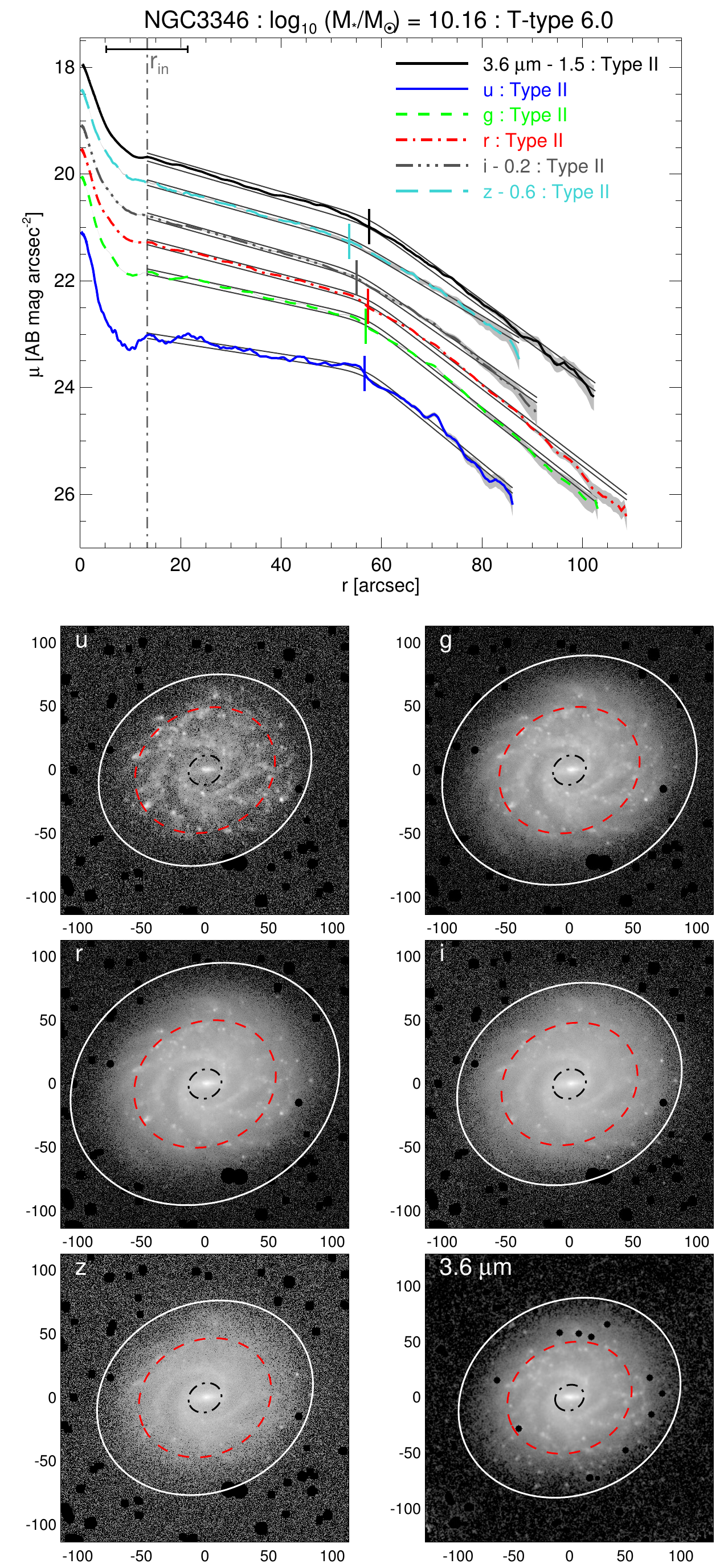}}	
    \end{minipage}%
    \begin{minipage}{0.46\textwidth}
        \resizebox{\hsize}{!}{\includegraphics{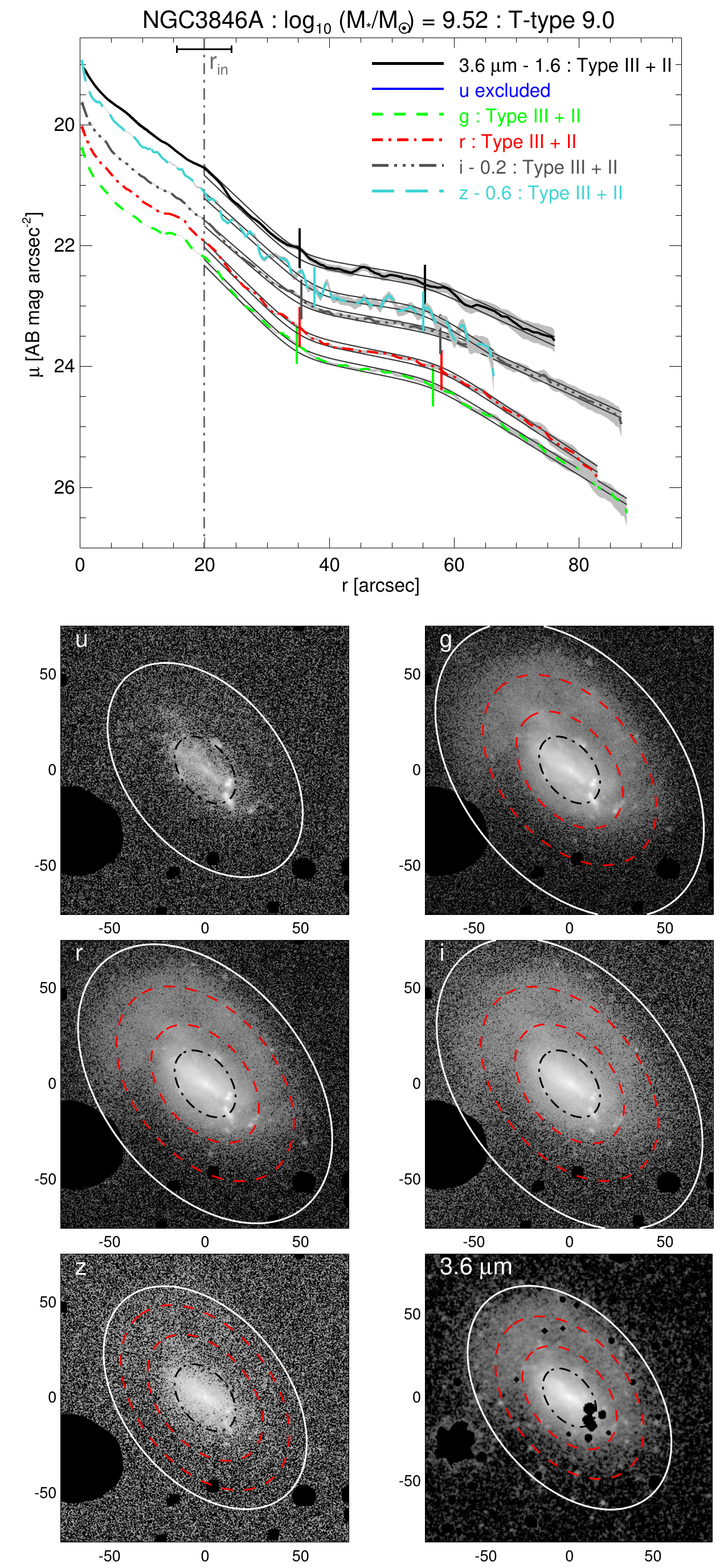}}	
    \end{minipage}%
    \caption{Examples of the radial surface brightness profiles and masked images of different bands, for a prototypical Type II break galaxy NGC 3346 (left panels), and more rare galaxy NGC 3846A with inner Type III and outer Type II break (right panels). The \emph{u}-band data was excluded from analysis for NGC 3846A, because the radial profile could not be traced to large enough radius. The profiles at \emph{i}, \emph{z} and $3.6 \, \mu m$ are offset with the values indicated in the legend in order to separate the profiles. The profiles have been cut by $r_{\rm out}$, while the $r_{\rm in}$ plotted here with vertical dot-dashed line is the same in all bands  (see Section \ref{fitting} for definitions). The grey area under each profile indicates the uncertainty due to over or under subtraction of the sky by $\sigma_{\rm sky}$. The break radius ($R_{\rm BR}$) in each band is plotted with a short vertical line segment, while the disc fits are plotted with hollow double-lines. In the images $r_{\rm in}$ is plotted with a black dot-dash line, $R_{\rm BR}$ with a red dashed line, and $r_{\rm out}$ with a solid white line. The images show surface brightness range $\mu=[17, 27] \, \text{AB mag arcsec}^{-2}$, and the x and y-axis scales are in arcseconds. Similar figures for all the galaxies are available at \url{http://www.oulu.fi/astronomy/DISC_BREAKS/}}
    \label{example-prof-ima}
\end{figure*}

The \emph{r}-band images of SDSS, or the \emph{g}-band images in case of the Liverpool telescope data, are used to create initial point source masks using SExtractor for the optical data \citep{bertin1996}. These masks are further manually edited so that they are suitable for the study of the radial surface brightness profiles in all the optical bands (see Fig. \ref{example-prof-ima} for examples of masked images). The use of different masks for infrared and optical data was motivated by the large differences in the amount of background sources in the images. The 3.6 $\mu m$ S$^4$G images have a large number of faint and small sources in the images, which are not present in the SDSS data. To examine the effect of the different masks, we acquired the disc profile parameters from the surface brightness profiles of $\sim 300$ galaxies using the same mask in every band, and found that the changes in the disc parameters are negligible. The galaxy centre coordinates of the S$^4$G images were converted to the optical image coordinates using the astrometry information in the image headers. Our checks indicated that the astrometry is accurate within $\approx 0.5$ arcseconds, comparable to $\approx 1-2$ pixels in the SDSS data. This accuracy is enough for our purposes, because large scale outer structures are studied. The optical images were background subtracted, but nevertheless the background sky level was checked and $\sigma_{\rm sky}$  was calculated for every image. This was done in order to determine to what surface brightness level the profiles can be reliably traced, which will be discussed in more detail in section \ref{fitting}.

\begin{table*}
\centering
\caption{Galaxy properties. }
\label{galaxy_properties}
\begin{tabular}{l l l l l l l l l l}
\hline
\hline
Galaxy & Survey & Bands & Hubble & Stellar mass                 & Distance & $R_{25}^{B}$  & $R_{24}^{3.6}$ & P.A.    & $b/a$ \\
       &        &       & T-type & [$\log_{10} (M_*/M_{\sun})$] & [Mpc]    & [arcsec]      & [arcsec]      & [\degr] & [$1-\epsilon$] \\
\hline
ESO013-016 & S$^4$G & $3.6\mu m$ & 6.0 & 9.6 & 22.0 $\pm$ 2.3 & 64.3 & 57.9 & -14.3 & 0.66 \\
ESO026-001 & S$^4$G & $3.6\mu m$ & 5.5 & 9.4 & 17.0 $\pm$ 7.9 & 45.4 & 45.3 & 19.3 & 0.94 \\
ESO027-001 & S$^4$G & $3.6\mu m$ & 3.5 & 10.0 & 15.1 $\pm$ 11.4 & 81.3 & 86.6 & 12.5 & 0.78 \\
ESO079-005 & S$^4$G & $3.6\mu m$ & 9.0 & 9.3 & 21.2 $\pm$ 3.2 & 58.4 & 43.0 & -8.8 & 0.59 \\
ESO079-007 & S$^4$G & $3.6\mu m$ & 8.0 & 9.5 & 25.1 $\pm$ 3.7 & 48.2 & 41.0 & 18.3 & 0.81 \\
ESO085-047 & S$^4$G & $3.6\mu m$ & 9.0 & 8.5 & 16.7 $\pm$ NA & 34.8 & 22.5 & 31.6 & 0.54 \\
ESO137-010 & NIRS0S & $K_{\rm s}$ & -1.0 & 11.0 & 43.0 $\pm$ 3.0 & 112.2 & 83.6 & 148.8 & 0.67 \\
NGC3846A & S$^4$G & $3.6\mu m$, \emph{u}, \emph{g}, \emph{r}, \emph{i}, \emph{z} & 9.0 & 9.5 & 25.3 $\pm$ NA & 58.1 & 44.0 & 38.4 & 0.65 \\
UGC12846 & S$^4$G & $3.6\mu m$, \emph{u}, \emph{g}, \emph{r}, \emph{i}, \emph{z} & 9.0 & 8.7 & 27.3 $\pm$ 3.2 & 46.3 & 13.9 & -4.7 & 0.87 \\
\hline
\end{tabular}
\tablefoot{Sample of the table which is available at the CDS. Data for each galaxy includes available bands, the mid-IR Hubble type \citep{buta2015,laurikainen2011}, stellar mass \citep{munozmateos2015}, redshift-independent distance and the standard deviation from NED (NA indicates that standard deviation is not available, if redshift-independent distance was not available the values were calculated from the redshifts obtained from NED with $H_0 = 72$ km/s / Mpc), 25 mag arcsec$^{-2}$ radius in the B-band from HyperLeda ($R_{25}^{B}$), the 24 mag arcsec$^{-2}$ radius of the infrared data ($R_{24}^{3.6}$), and position angle and minor/major axis ratio of the outer disc as measured from the infrared data \citep{laurikainen2011,salo2015}.}
\end{table*}


\subsection{Obtaining the surface brightness profiles}

The surface brightness profiles used to study the galaxy discs were created by running the IRAF\footnote{IRAF is distributed by the National Optical Astronomy Observatories, which are operated by the Association of Universities for Research in Astronomy, Inc., under cooperative agreement with the National Science Foundation.} task \emph{ellipse}, with the centre, ellipticity, and position angle fixed to the values of the outer discs. Surface brightness profiles of the S$^4$G galaxies were converted to AB-system magnitudes from the pixel values $F_i$ (MJy str$^{-1}$) in the following manner:
\begin{equation}
  \mu = -2.5 \, \log_{10} \left( F_i \right) + 20.472,
\end{equation}
where the value 20.472 is the surface brightness zero-point of the flux to magnitude conversion, as calculated from the definition of the AB magnitude scale \citep{oke1974}. 

The flux calibration of the NIRS0S images was based on the 2MASS Vega system. The median difference in the aperture magnitudes between early-type NIRS0S galaxies ($-3 \leq T \leq 0$) and S$^4$G is $2.566 \pm 0.082$ \citep{laine2014}. This value was used to convert the magnitudes from the Vega to $3.6 \, \mu m$ AB-system, and it is added when constructing the surface brightness profiles of the NIRS0S galaxies:
\begin{equation}
  \mu = -2.5 \, \log_{10} \left( ADU \right) + 2.566.
  \label{nirsosconv}
\end{equation}
The above conversion has been applied to the surface brightness values of the $K_{\rm s}$-band data of NIRS0S galaxies given in Table \ref{break_parameters}, and in all the figures of this study.

The optical images provided by \citet{knapen2014} have headers which include the magnitude zeropoints ($\mathrm{mag}_{\rm zp}$) and the pixel scales ($\mathrm{pix}$). The surface brightness profiles are then constructed in the following manner:
\begin{equation}
	\mu = -2.5 \, \log_{10} \left(ADU \right) + \mathrm{mag}_{\rm zp} + 5 \log_{10} \left( \mathrm{pix} \right).
	\label{surf-eq-opt}
\end{equation}

\begin{table}
\centering
\caption{The limiting surface brightness and outer radius of profiles.}
\begin{tabular}{l l l l l} 
\hline
\hline
Band & $\mu(r_{\rm out})$         & $r_{\rm out} / R_{25}^B$ & N$_{\rm bad}$ & N$_{\rm good}$  \\
     & [AB mag arcsec$^{-2}$] &                    & \\
\hline
\emph{u}              & $26.1 \pm 0.6$ & $1.1 \pm 0.3$ & 127 & 323  \\
\emph{g}              & $26.2 \pm 0.6$ & $1.5 \pm 0.4$ & 15  & 465 \\
\emph{r}              & $25.8 \pm 0.6$ & $1.5 \pm 0.4$ & 2   & 448 \\
\emph{i}              & $25.0 \pm 0.6$ & $1.4 \pm 0.4$ & 15  & 435 \\
\emph{z}              & $24.0 \pm 0.6$ & $1.1 \pm 0.3$ & 111 & 339 \\
\emph{K}$_{\rm s}$    & $22.5 \pm 0.5$ & $0.9 \pm 0.2$ & $\ldots$ & 28  \\
$3.6 \mu m$           & $25.1 \pm 0.4$ & $1.3 \pm 0.4$ & $\ldots$   & 725 \\
\hline \\
\end{tabular}
\tablefoot{Median values of the surface brightness at the outer limit of the fit range ($r_{\rm out}$, see Section \ref{fitting} for definition) in different bands, and the median value of the fit range limit relative to the B-band 25  mag arcsec$^{-2}$ radius ($R_{25}^B$, taken from HyperLeda \citep{makarov2014}). The value N$_{\rm bad}$ indicates the number of galaxies with excluded surface brightness profiles in each band (see Section \ref{fitting}), while the value N$_{\rm good}$ indicates the number of galaxies with usable surface brightness profiles.}
\label{table:sigmasky}
\end{table}

\subsection{Profile function fitting}
\label{fitting}

We use the method by \citet{laine2014} to fit the different functions to the surface brightness profiles of the galaxies. The used function depends on the shape of the surface brightness profile, determined by visual inspection. Detailed description of the fitting method and the different functions used can be found in Section 3.3 of \cite{laine2014}. Similar methods have been used also in some other studies of disc breaks (e.g. \citealt{erwin2005,pohlen2006,erwin2008,munozmateos2013}). 

We first examined the surface brightness profiles of the infrared data in the same manner as in \cite{laine2014}. We ignored the inner regions of the galaxies, dominated by bulges and bars, and fit functions to the surface brightness profile where the disc light dominates. The outermost radius of the fit ($r_{\rm out}$) is defined by the radius where over- or underestimating the background sky level by $\sigma_{\rm sky}$ affects the surface brightness profile more than $\pm 0.2$ mag arcsec$^{-2}$. The inner fit radius ($r_{\rm in}$) is always selected manually. Inspecting the galaxy images we checked that this radius is outside bars and bulges. An example of how the fit range is selected is shown in Figure \ref{example-prof-ima}. 

Then we automatically fitted the data in all the optical bands using the same functions and the same initial parameter values as in the infrared data. The inner limit of the fit range is the same in all bands for the individual galaxies. As previously noted, the outer limit depends on $\sigma_{\rm sky}$, and it is calculated independently for each band. Each profile and the fit was visually checked and the fit function changed if the profile type was different in some band (e.g. Type II in $3.6 \, \mu m$ infrared data and Type I in \emph{r}-band). If the data in some band is significantly less deep we marked the profile and the fit as `bad', and removed it from further analyses. These are the cases where the image depth is not sufficient for a reliable determination of the disc parameters. The depth of the `good' data, and the number of good and bad profiles, are shown in Table \ref{table:sigmasky}. In general, the data of the 3.6 $\mu m$, \emph{g}, \emph{r}, and \emph{i} bands are comparable in depth, but the \emph{u}, \emph{z}, and NIRS0S $K_{\rm s}$ band data are shallower. The image depth achieved here is similar with the previous studies of disc breaks using the same data sources (e.g. SDSS data in \citealt{pohlen2006,erwin2008}, and S$^4$G data used in \citealt{martinnavarro2012,munozmateos2013}). 
		
To estimate the uncertainties of the disc and break parameters we used a Monte Carlo approach (see also \citealt{munozmateos2013,laine2014}). We assumed that the largest source of uncertainty is due to the selection of the fitted region, rather than the statistical error of the fit to the data. We alter the disc-fit region-limiters independently by drawing 500 new values for $r_{\rm in}$ and $r_{\rm out}$, starting from a uniform distribution, which is centred on the original limit value and has a width of $0.15 \times R_{24}^{3.6}$, measured from the infrared data (using Eq. \ref{nirsosconv} to convert from $K_{\rm s}$ to 3.6 $\mu m$ in the case of NIRS0S galaxies). For example, for NGC 3346 in Fig. \ref{example-prof-ima} this equals to 16.3\arcsec wide distribution. The $R_{24}^{3.6}$ radius is chosen because it is reached also by the shallower NIRSOS data. The profile was then fitted with these new limits, and the standard deviation of the different parameters returned was calculated. The profile type was kept the same, and the initial values for the fits are also the same in this process. The disc and break parameter values reported in the following sections and figures are from the initial fit, and the uncertainties retrieved with the above methods are included in Table \ref{break_parameters}, which is available in full form at the CDS.

\subsection{Break type classification}
\label{break_types}

Following the main classification criteria \citep{pohlen2006,erwin2008}, the profiles are divided into four categories by the behaviour of the outer disc:
\begin{itemize}
    \item Type I: single exponential.
    \item Type II.i: a break is seen inside or at the bar radius, and the outer disc is single exponential.
    \item Type II: down-bending break in the outer disc.
    \item Type III: up-bending break in the outer disc.
\end{itemize}
Some galaxies have even more complex disc profiles with more than one break in the outer discs, showing a combination of Type II and III breaks (an example is shown in Fig. \ref{example-prof-ima}). A number of previous studies have further divided the Type II breaks into several different subgroups (e.g. \citealt{pohlen2006,erwin2008,gutierrez2011}), which are not used here. However, we used a comparable approach and systematically associate the Type II break location in the infrared data with different structural components of the galaxies, in a similar manner as in \cite{laine2014}.  In practice, we connected the breaks with: 
\begin{itemize}
    \item inner rings or lenses (r/l in Table \ref{break_parameters}),
    \item outer rings or lenses (R/L), 
    \item radial locations of bright star forming regions in the discs (BS),
    \item visual outer edges of the spiral structure (S),
    \item clearly asymmetric outer discs (AB).
\end{itemize}
The breaks associated with rings or lenses (r/l/R/L), and those associated with spiral structures (BS/S) form two major groups of the Type II breaks (groups a and b, respectively), which will be discussed in more detail in section \ref{struct_connection}. Cases in which we cannot find any association with a particular galaxy structure component we mark as NA in Table \ref{break_parameters}.

\begin{table*}
\centering
\caption{Surface brightness profile parameters.} 
{\tiny
\begin{tabular}{l l l l l l l l l l l l l l l l}
\hline	
\hline
Galaxy & Band & Profile & $r_{\rm in}$ & $r_{\rm out}$ & $\mu_{r_{\rm out}}$ & $R_{\rm BR}$  & $\mu_{\rm BR}$             & $\mu_{\circ \rm ,i}$          & $h_{\rm i}$     & $\mu_{\circ \rm, o}$          & $h_{\rm o}$  \\
       &      & type & [\arcsec] & [\arcsec] & [AB $\frac{\rm mag}{\rm arcsec^{2}}$ ]& [\arcsec] & [AB $\frac{\rm mag}{\rm arcsec^{2}}$] & [AB $\frac{\rm mag}{\rm arcsec^{2}}$] & [\arcsec] & [AB $\frac{\rm mag}{\rm arcsec^{2}}$] & [\arcsec] \\
\hline
ESO013-016 &  $3.6 \, \mu m$  & II (S)  & 11.5 & 70.6 & 25.4 & 47.0 $\pm$ 0.8 & 23.76 $\pm$ 0.05 & 21.50 $\pm$ 0.02 & 22.9 $\pm$ 0.3 & 20.57 $\pm$ 0.05 & 16.1 $\pm$ 0.2 \\
ESO026-001 &  $3.6 \, \mu m$  & II (R)  & 10.7 & 56.0 & 25.2 & 33.5 $\pm$ 0.6 & 23.03 $\pm$ 0.03 & 21.88 $\pm$ 0.05 & 35.2 $\pm$ 2.4 & 19.64 $\pm$ 0.07 & 11.1 $\pm$ 0.2 \\
ESO027-001 &  $3.6 \, \mu m$  & I & 29.5 & 95.1 & 24.9 & $\ldots$ & $\ldots$ & 20.08 $\pm$ 0.05 & 21.5 $\pm$ 0.3 & $\ldots$ & $\ldots$ \\
ESO079-005 &  $3.6 \, \mu m$  & I & 21.7 & 51.7 & 24.9 & $\ldots$ & $\ldots$ & 21.25 $\pm$ 0.04 & 14.1 $\pm$ 0.2 & $\ldots$ & $\ldots$ \\
ESO079-007 &  $3.6 \, \mu m$  & I & 4.6 & 53.7 & 25.4 & $\ldots$ & $\ldots$ & 21.10 $\pm$ 0.03 & 14.1 $\pm$ 0.2 & $\ldots$ & $\ldots$ \\
ESO085-047 &  $3.6 \, \mu m$  & III & 1.9 & 67.6 & 26.1 & 31.1 $\pm$ 0.5 & 25.15 $\pm$ 0.02 & 23.16 $\pm$ 0.02 & 16.2 $\pm$ 0.3 & 24.51 $\pm$ 0.02 & 46.7 $\pm$ 0.6 \\
ESO137-010 &  $K_{\rm s}$  & I & 27.3 & 71.5 & 24.6 & $\ldots$ & $\ldots$ & 20.79 $\pm$ 0.01 & 20.6 $\pm$ 0.0 & $\ldots$ & $\ldots$ \\
NGC3846A &  \emph{u}  & Ex. & $\ldots$ & $\ldots$ & $\ldots$ & $\ldots$ & $\ldots$ & $\ldots$ & $\ldots$ & $\ldots$ & $\ldots$ \\
NGC3846A &  \emph{g}  & III+II & 19.9 & 87.7 & 26.8 & 34.8 $\pm$ 3.0 & 22.28 $\pm$ 0.20 & 20.69 $\pm$ 0.00 & 10.5 $\pm$ 0.0 & 23.89 $\pm$ 0.03 & 53.3 $\pm$ 0.0 \\
NGC3846A &  \emph{g}  &   & 19.9 & 87.7 & 26.8 & 56.6 $\pm$ 3.2 & 22.96 $\pm$ 0.15 & 23.89 $\pm$ 0.03 & 53.3 $\pm$ 0.0 & 21.15 $\pm$ 0.02 & 17.1 $\pm$ 0.0 \\
NGC3846A &  \emph{r}  & III+II & 19.9 & 83.0 & 26.3 & 35.2 $\pm$ 3.2 & 21.94 $\pm$ 0.21 & 20.42 $\pm$ 0.00 & 10.7 $\pm$ 0.0 & 23.48 $\pm$ 0.03 & 53.5 $\pm$ 0.0 \\
NGC3846A &  \emph{r}  &   & 19.9 & 83.0 & 26.3 & 57.9 $\pm$ 3.2 & 22.75 $\pm$ 0.16 & 23.48 $\pm$ 0.03 & 53.5 $\pm$ 0.0 & 20.55 $\pm$ 0.03 & 16.0 $\pm$ 0.0 \\
NGC3846A &  \emph{i}  & III+II & 19.9 & 86.8 & 25.6 & 35.5 $\pm$ 3.1 & 21.77 $\pm$ 0.18 & 20.51 $\pm$ 0.00 & 12.1 $\pm$ 0.0 & 23.21 $\pm$ 0.03 & 60.9 $\pm$ 0.0 \\
NGC3846A &  \emph{i}  &   & 19.9 & 86.8 & 25.6 & 57.8 $\pm$ 3.2 & 22.56 $\pm$ 0.12 & 23.21 $\pm$ 0.03 & 60.9 $\pm$ 0.0 & 21.36 $\pm$ 0.01 & 22.9 $\pm$ 0.0 \\
NGC3846A &  \emph{z}  & III+II & 19.9 & 66.3 & 25.1 & 37.6 $\pm$ 3.1 & 21.74 $\pm$ 0.23 & 20.39 $\pm$ 0.00 & 11.7 $\pm$ 0.0 & 23.47 $\pm$ 0.04 & 82.3 $\pm$ 0.0 \\
NGC3846A &  \emph{z}  &   & 19.9 & 66.3 & 25.1 & 55.1 $\pm$ 3.1 & 22.35 $\pm$ 0.19 & 23.47 $\pm$ 0.04 & 82.3 $\pm$ 0.0 & 20.51 $\pm$ 0.04 & 16.5 $\pm$ 0.0 \\
NGC3846A &  $3.6 \, \mu m$  & III+II (AB)  & 19.9 & 73.5 & 25.5 & 35.2 $\pm$ 0.5 & 24.12 $\pm$ 0.05 & 20.88 $\pm$ 0.09 & 11.1 $\pm$ 0.4 & 23.87 $\pm$ 0.06 & 67.4 $\pm$ 7.3 \\
NGC3846A &  $3.6 \, \mu m$  &   & 19.9 & 73.5 & 25.5 & 55.3 $\pm$ 0.4 & 24.73 $\pm$ 0.02 & 23.87 $\pm$ 0.06 & 67.4 $\pm$ 7.3 & 22.01 $\pm$ 0.03 & 22.6 $\pm$ 0.2 \\
UGC12846 &  \emph{u}  & Ex. & $\ldots$ & $\ldots$ & $\ldots$ & $\ldots$ & $\ldots$ & $\ldots$ & $\ldots$ & $\ldots$ & $\ldots$ \\
UGC12846 &  \emph{g}  & I & 10.6 & 44.5 & 26.6 & $\ldots$ & $\ldots$ & 23.02 $\pm$ 0.02 & 15.7 $\pm$ 0.2 & $\ldots$ & $\ldots$ \\
UGC12846 &  \emph{r}  & I & 10.6 & 44.7 & 26.0 & $\ldots$ & $\ldots$ & 22.61 $\pm$ 0.01 & 14.5 $\pm$ 0.1 & $\ldots$ & $\ldots$ \\
UGC12846 &  \emph{i}  & I & 10.6 & 37.9 & 25.1 & $\ldots$ & $\ldots$ & 22.48 $\pm$ 0.02 & 15.9 $\pm$ 0.2 & $\ldots$ & $\ldots$ \\
UGC12846 &  \emph{z}  & I & 10.6 & 32.0 & 24.7 & $\ldots$ & $\ldots$ & 22.28 $\pm$ 0.06 & 15.6 $\pm$ 0.6 & $\ldots$ & $\ldots$ \\
UGC12846 &  $3.6 \, \mu m$  & I & 10.6 & 50.5 & 26.5 & $\ldots$ & $\ldots$ & 23.29 $\pm$ 0.02 & 16.4 $\pm$ 0.1 & $\ldots$ & $\ldots$ \\
\hline 
\end{tabular} }
\tablefoot{Sample of the table which is available at the CDS. From the disc profile fits we first give the filter and profile type (Ex. = excluded from analysis). In parenthesis we give the galaxy structure with which the Type II break has been visually connected with in the infrared data: NA = no visual association with a specific galaxy structure, BS = in spiral arms near prominent star formation regions, S = visual spiral outer edge, r/R = inner-/outer-ring, l/L = inner-/outer-lens, AB = asymmetric outer disc. The inner ($r_{\rm in}$) and outer radius ($r_{\rm out}$) of the fit region are also given, as well as the surface brightness at $r_{\rm out}$ ($\mu_{r_{\rm out}}$). The break radius ($R_{\rm BR}$), surface brightness at the break ($\mu_{\rm BR}$), and disc central surface brightness and scalelengths inside ($\mu_{\circ \rm ,i}$, $h_{\rm i}$) and outside ($\mu_{\circ \rm ,o}$, $h_{\rm o}$) the break are also given. The values of $\mu$ are inclination corrected ($\mu=\mu_{\rm org}-2.5\log_{10}(b/a)$), but are not corrected for extinction by Galactic dust. Note that the $K_{\rm s}$ band surface brightness values have been converted to 3.6 $\mu m$ AB-system using Eq. \ref{nirsosconv}. }
\label{break_parameters}
\end{table*}

In \cite{pohlen2006}, cases in which the Type II break arises from asymmetry of the outer disc (Type II-AB) were discarded from statistics. They were not considered as true features of the disc.  In this study the clearly asymmetric cases are included into Type II. This is the case for example in NGC 3846A, shown in Figure \ref{example-prof-ima}. Estimating  in a quantitative manner whether asymmetric disc causes the observed break has turned out to be difficult. We tested whether the asymmetry parameter $A$ measured for the S$^4$G galaxies in the 3.6 $\mu m$ data by \cite{holwerda2014} could be used for that purpose. This parameter is defined as:
\begin{equation}
    A = \frac{\Sigma_{i,j} | I(i,j) - I_{180}(i,j)|}{2 \Sigma_{i,j}|I(i,j)|},
\end{equation}
where $I(i,j)$ is the pixel intensity in position (i,j), and $I_{180}(i,j)$ is the pixel intensity in the same position in an image rotated by $180^{\circ}$. In completely symmetric galaxies these terms cancel each other, and $A$ would be very small. However, the application of $A$ to separate the Type II-AB cases turned out to be difficult due to the large scatter of this parameter: for example, the fairly symmetric Type II galaxy NGC 3346 in Figure \ref{example-prof-ima} (left panels) has $A=0.52$, whereas NGC 3846A (right panels) has a lower value of $A=0.45$, although by eye it appears clearly more asymmetric. The classification of the Type II-AB cases would have to be done subjectively even with the aid of this quantitative measure of asymmetry.

Type III breaks have been previously divided into two subgroups based on the behaviour of the ellipticity in the outer disc (e.g. \citealt{erwin2005}). When the ellipticity decreases in the outer edges of the galaxies the break is considered as an effect of a galactic halo or a large spheroidal (Type III-s). In cases where this behaviour is not seen, or spiral structure  continues beyond the break radius, the break has been attributed to the disc (Type III-d). However, in nearly face-on galaxies this division cannot be made (the discs themselves have nearly circular geometry), for which reason  we did not use these subgroups here.

The division of the Types II and III to various subgroups relies mainly on subjective criteria, which is an general problem of the disc break studies. There is no consensus even on how large the deviations from a single exponential profile should be for a galaxy to have a Type II or III break. In Appendix A we show that this ambiguity causes some differences between disc break studies even when using the same methodology.


\begin{table*}
\centering
\caption{Statistics of the break types in the surface brightness profiles of the infrared data. }
\begin{tabular}{l l l l l l l l l}
\hline
\hline
\multicolumn{3}{l}{All galaxies (N=753)} & \multicolumn{3}{l}{Low mass bin (N=372)} &\multicolumn{3}{l}{High mass bin  (N=381)}  \\
Type & Fraction & Bar fraction & Type & Fraction & Bar fraction & Type & Fraction & Bar fraction \\
\hline
I    & $30 \pm 2 \%$ (228) & $73 \pm 3 \%$  & I    & $35 \pm 2 \%$ (131) & $73 \pm 4 \%$  & I    & $25 \pm 2 \%$  (97)  & $72  \pm 5 \%$ \\
II.i & $6  \pm 1 \%$ (43)  & $100 \pm 0 \%$ & II.i & $6  \pm 1 \%$ (21)  & $100 \pm 0 \%$ & II.i & $6 \pm 1 \%$   (22)  & $100  \pm 0 \%$ \\
II   & $44 \pm 2 \%$ (335) & $72 \pm 2 \%$  & II   & $38 \pm 3 \%$ (143) & $66 \pm 4 \%$  & II   & $50 \pm 3 \%$ (192) & $76  \pm 3 \%$ \\
III  & $25 \pm 2 \%$ (189) & $51 \pm 4 \%$  & III  & $26 \pm 2 \%$ (98)  & $50 \pm 5 \%$  & III  & $24 \pm 2 \%$  (91) & $53  \pm 5 \%$ \\
\hline
\end{tabular}
\label{table:infra-stat}
\tablefoot{The profile types are divided to low and high mass bins by the median mass in the Hubble type bins as shown in Figure \ref{sample_histo}. We also give the bar fractions for each profile type. The uncertainties are calculated using binomial statistics, and denote the $\pm 1 \sigma$ uncertainties. Some galaxies show two breaks in their discs and are counted twice in the statistics, explaining why the overall percentages can be above 100\%.}
\end{table*}

\section{Results}
\label{results}

\subsection{Break-type distributions}
\label{results:wave-mass}

\begin{figure*}
  \begin{center}
    \includegraphics[width=\textwidth]{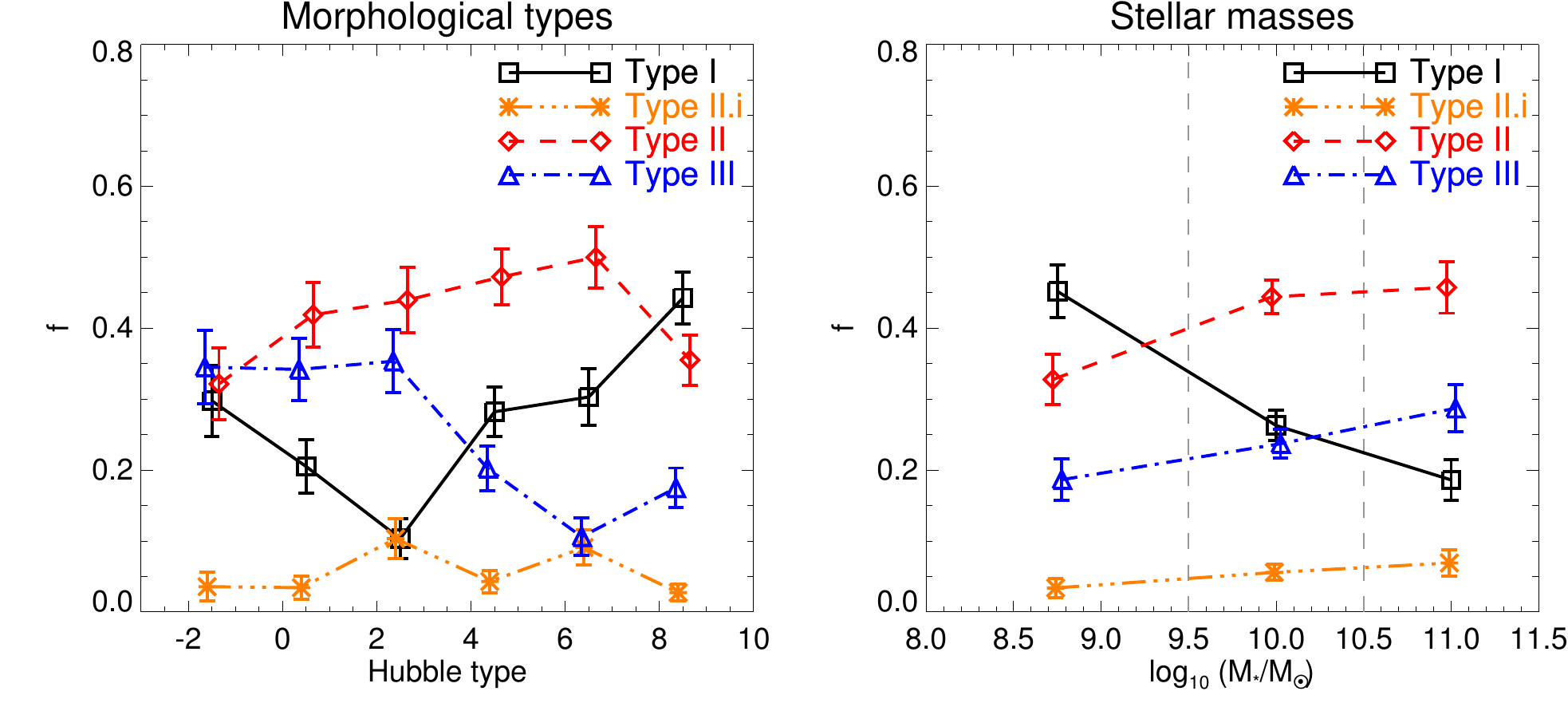}
    \caption{Fractions of the break types in the infrared sample, as a function of the Hubble type, and galaxy stellar mass. The error bars are calculated using binomial statistics, and denote the $\pm 1 \sigma$ uncertainties. In the right panel the vertical grey dashed lines indicate the bin limits, selected so that each bin has roughly equal number of galaxies.}
    \label{ir_sample_types}
      \end{center}
\end{figure*}

The wide coverage of stellar masses enables a study of the profile type fractions as a function of stellar mass of the galaxies. In Figure \ref{ir_sample_types} (right panel) we show that the fraction of single exponential discs (Type I) decreases, while the fraction of Type II and III breaks increases with increasing stellar mass. The Type II.i profiles are only marginally more common in more massive galaxies. These trends are visible also in Hubble types  (Fig. \ref{ir_sample_types} left panel), because late type galaxies are generally less massive (see Fig. \ref{sample_histo}). Within the uncertainties the obtained fractions in the infrared (see Table \ref{table:infra-stat} leftmost column) are similar to those found previously in \citet{laine2014} for bright galaxies, which form part of the current sample. As expected, for the bright galaxies the distribution of the profile types among the Hubble types (Fig. \ref{ir_sample_types} left panel) is similar to that found by \cite{laine2014} (Fig. 5. upper panel in their study). 

\begin{figure*}   
	   \begin{center}
    \includegraphics[width=\textwidth]{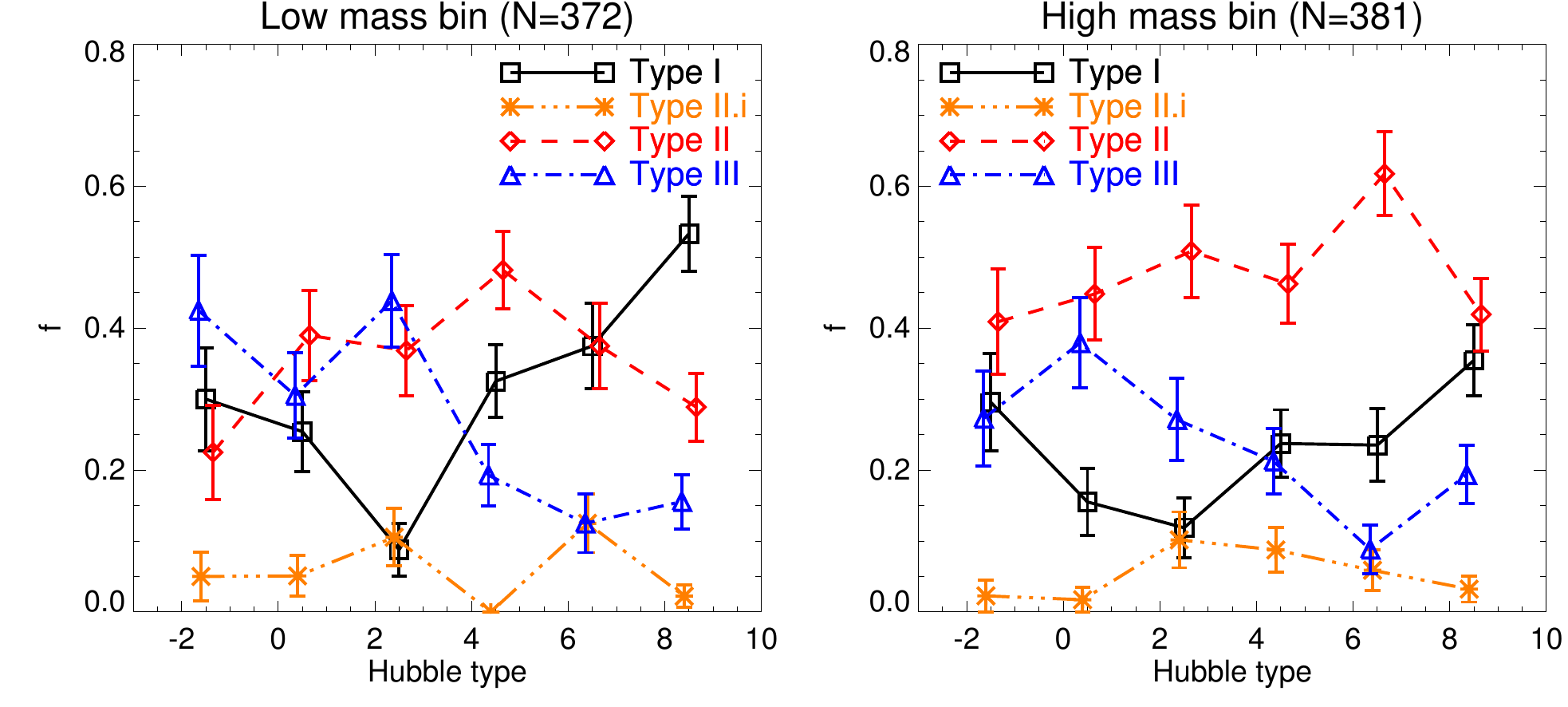}
    \caption{Fractions of the break types as a function of the galaxy Hubble type in the infrared sample shown separately for low mass (\emph{left panel}), and high mass bins (\emph{right panel}). The galaxies are divided to low and high mass bins by the median mass of the Hubble type bins shown in Figure \ref{sample_histo}. The error bars are calculated using binomial statistics, and denote the $\pm 1 \sigma$ uncertainties. }
    \label{ir_sample_types_highlow}
  \end{center}
\end{figure*}

We further separated the infrared sample into low and high mass bins, using the median mass in each Hubble type bin as shown in Figure \ref{sample_histo}. As discussed above, the fraction of Type I profiles changes significantly with stellar mass, being $35\%$ in the low-mass bin (see Table \ref{table:infra-stat} middle column), and $25\%$ in the high-mass bin (Table \ref{table:infra-stat} right column). The fraction of Type II profiles increases from $38\%$ in low-mass bin to $50\%$ in high-mass bin, whereas Type III fraction stays the same. The distributions of the fractions of the profile types for low and high-mass bins are similar among all Hubble types (Fig. \ref{ir_sample_types_highlow}). The Type III profiles are most common in the early-type galaxies ($T \lesssim 4$), whereas the type I fraction increases towards late-type galaxies. In the high-mass bin the Type II profiles are the most common profile type in all Hubble-type bins (see Fig. \ref{ir_sample_types_highlow}, right panel).

\citet{pohlen2006}, \citet{erwin2008}, and \citet{gutierrez2011} have previously studied the surface brightness profile breaks in the optical \emph{g}, \emph{r}, \emph{R}, and \emph{B} bands  for bright S0-Sdm galaxies  (median $M_{B} = -19.6$, corresponds roughly to $\log_{10} (M_{*}/M_{\sun}) \approx 10.2$, see Fig. \ref{other_studies}). The fractions of the break types in Table \ref{table:infra-stat} for the high-mass bin agree well with the fractions reported in \cite{gutierrez2011} for Type I and II profiles ($21 \pm 3\%$ and $50 \pm 4 \%$, respectively in \cite{gutierrez2011}, and $25 \pm 2 \%$ and $50 \pm 3 \%$ here). \cite{gutierrez2011} find a fraction of $38 \pm 4\%$ for Type III profiles, compared to $24 \pm 2 \%$ that we find for the high-mass bin. However, the galaxy stellar mass limits are different in the two studies which could explain the difference.

\citet{herrmann2013} have studied the disc breaks in dwarf galaxies using ultraviolet, optical, and infrared observations. Their sample consists mostly of dwarf irregular galaxies (96 out of 141, 26 blue compact dwarfs, and 19 Magellanic spirals). They find that $61\%$ of the galaxies in their sample at all bands show a Type II profile. Only $8\%$ had Type I profiles, $16\%$ Type III profiles, and $6\%$ had a combination of Type II and III profiles. These results for Types I and II are significantly different when compared to our results in the lowest mass bin in Figure \ref{ir_sample_types} (right panel), or with the fractions given in Table \ref{table:infra-stat} (middle column). \cite{herrmann2013} shows that irregular dwarf galaxies almost always have a Type II break. Our sample has only a small number of such small-mass galaxies (see Fig. \ref{other_studies}), but we will briefly address this difference in Section \ref{discuss:mass}.


We find that the profile type is generally independent of the observed wavelength band if the surface brightness profiles can be reliably traced to a similar radius (see also \citealt{macarthur2003} and \citealt{herrmann2013}). Based on the median level to which the surface brightness can be followed (see Table \ref{table:sigmasky}), the bands \emph{g}, \emph{r}, \emph{i} and $3.6 \, \mu m$ are similar in depth. In these bands the profile type was found to be the same in 92 \% of the galaxies (400 out of 433). Furthermore, in cases in which the profiles could be reliably studied in all six bands, the profile type was found to be the same in 90\% of the cases (258 out of 286). This is a similar fraction to that found by \cite{herrmann2013} for the low-mass galaxies at the end of the Hubble sequence, by comparing the properties of the discs in ultraviolet, optical, and infrared bands.

In most of the galaxies in which the profile type depends on wavelength band, the profiles are single exponentials in the infrared, and have either Type II (sometimes also Type II.i) or Type III in \emph{g}, \emph{r}, and \emph{i} bands.  In some galaxies an additional second outer break appears in the \emph{g}, \emph{r}, or \emph{i} bands. These differences actually arise from the ability to follow the surface brightness profiles to slightly larger radii in these bands compared to the infared data, allowing more reliable determination of the breaks at very large radii (17 out of the 33 cases in which the profile classification is different, see also Table \ref{table:sigmasky}). Only in very rare cases the break type changes from Type II to III, or vice versa, between the bands (e.g. four galaxies in the \emph{g} band). In many cases the \emph{u} and \emph{z} bands were excluded from further analysis, because the data was not deep enough to reach the break radius seen in other bands. This artificially increases the fraction of Type I profiles with respect to Types II and III in the \emph{u} and \emph{z} bands.

\begin{figure*}
  \begin{center}
    \includegraphics[width=\textwidth]{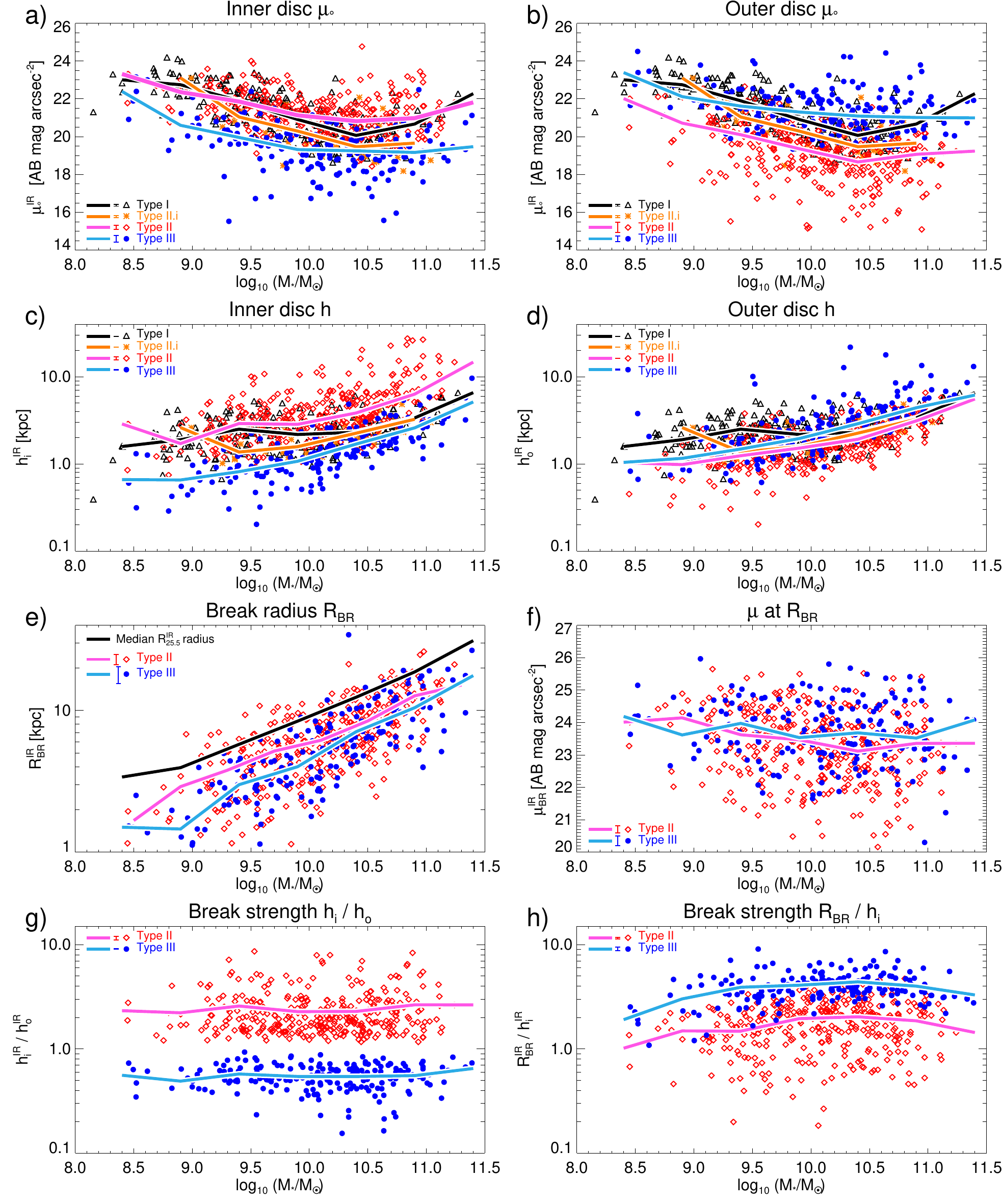}  
    \caption{Parameters of the discs and breaks in the infrared as a function of stellar mass. In every panel the solid lines indicates the running medians of the values. The surface brightness values have been corrected for inclination. The error bars near the figure legend indicate the mean errors of the values, calculated with propagation of errors. }
    \label{ii_iii_trad}
  \end{center}
\end{figure*}

\subsection{Parameters of the discs and breaks in the infrared as a function of galaxy stellar mass}

\subsubsection{Central surface brightness and scalelength of the disc}

We find that the scalelengths and the extrapolated central surface brightnesses of the discs correlate with galaxy stellar masses, which is consistent with that found in the previous studies (e.g. \citealt{courteau2007,gadotti2009,munozmateos2013}). In Figure \ref{ii_iii_trad} (panels a-d)  we show that these correlations hold independently also for the inner and outer disc sections of Type II and III profiles (see also \citealt{munozmateos2013}).  

The difference in the central surface brightness between the inner and outer disc ($|\mu_{\circ \rm ,i}^{IR}-\mu_{\circ \rm ,o}^{IR}|$) is on average $\sim 2$ mag arcsec$^{-2}$, which is very similar for Type II and III profiles ($2.1 \pm 1.2$ and $2.0 \pm 1.1$, respectively). Furthermore, 97 \% of Type II profiles have $|\mu_{\circ \rm ,i}^{IR}-\mu_{\circ \rm ,o}^{IR}| \sim$ 0.5--6, which is in agreement with \cite{munozmateos2013}, and with the predictions of the simulation models by  \cite{debattista2006}. Furthermore, 97 \% of Type III profiles have $|\mu_{\circ \rm ,i}^{IR}-\mu_{\circ \rm ,o}^{IR}| \sim$ 0.5--5. For Type I profiles the median values of $\mu_{\circ}^{IR}$ and $h^{IR}$ are very similar to those obtained for the inner discs of Type II profiles, particularly at galaxy stellar masses of $\log_{10} (M_{*}/M_{\sun}) \lesssim 10$. Also, $h_{\rm o}^{IR}$ is similar for Type II and III profiles.

\begin{figure*}
    \begin{center}
        \includegraphics[width=\textwidth]{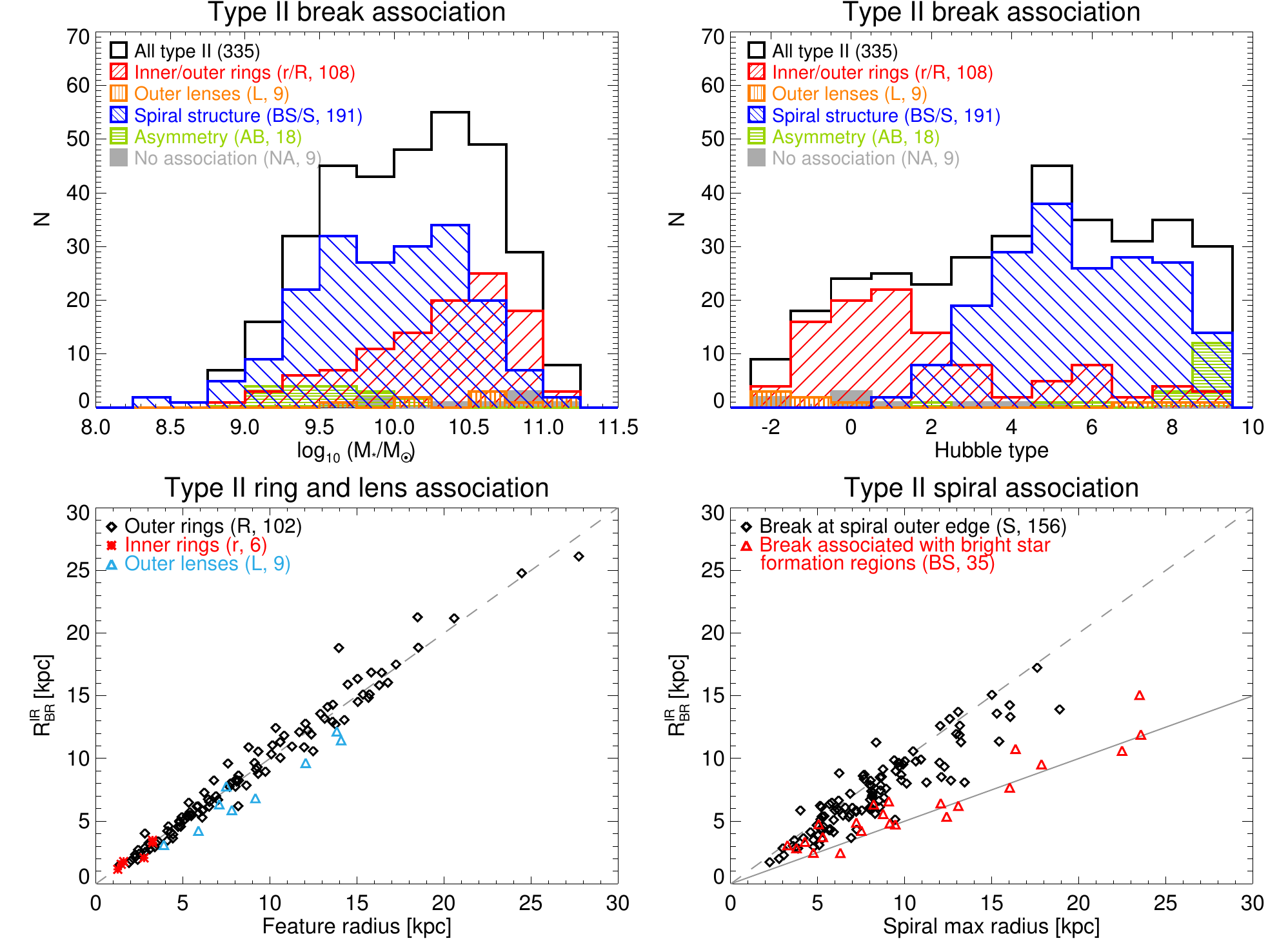}
        \caption{\emph{Upper left panel}: Distributions of Type II breaks associated with different structure components of the galaxies, as a function of stellar mass. \emph{Upper right panel}: Distribution among Hubble types. The Type II breaks connected to spirals in the \emph{upper panels} includes both the breaks seen at the spiral outer edges (S) and the breaks seen near prominent star formation regions inside the spirals (BS). \emph{Lower panels}: Sizes of the structures \citep{laurikainen2011,comeron2014,herreraendoqui2015} are correlated with the $R_{\rm BR}^{IR}$: in the left panel for rings and lenses, and in the right panel for spiral structures. The dashed line indicates where feature radius equals $R_{\rm BR}^{IR}$, and the solid line indicates where $R_{\rm BR}^{IR}$ equals  $0.5 \times$ feature radius. The letter in parenthesis is the same as the Type II break association given in Table 3, and the number given in parenthesis indicates the number of breaks included in each group.}
        \label{ii_mass_connection}
    \end{center}
\end{figure*}

\subsubsection{Break radius and break strength}

We find that in the infrared the break radius, $R_{\rm BR}^{IR}$, both in Type II and III profiles correlates with the galaxy stellar mass (Fig. \ref{ii_iii_trad} panel e), consistent with the previous studies both in optical and infrared (e.g. \citealt{pohlen2006,azzollini2008b,herrmann2013,munozmateos2013}). On average, Type III breaks appear at 0.5-1 kpc smaller radius than Type II breaks, although the scatter is large for both types. In galaxies with $M_{*} \sim 10^9 M_{\sun}$, $R_{\rm BR}^{IR} \sim 2-3$ kpc, while in galaxies with $M_{*} \sim 10^{10.5} M_{\sun}$, $R_{\rm BR}^{IR} \sim 10$ kpc. The correlation of $R_{\rm BR}^{IR}$ with stellar mass is very similar to the correlation between galaxy size and stellar mass, as shown with the running median of $R_{25.5}$ at $3.6 \, \mu m$. As expected, the surface brightnesses where the breaks appear, are slightly lower for the lower mass galaxies (Fig. \ref{ii_iii_trad} panel f). Majority ($\sim90\%$) of the breaks appear at the surface brightness interval of $22-25$ mag arcsec$^{-2}$ in 3.6 $\mu m$. \cite{herrmann2013} has showed also that the surface brightness in the optical V-band at $R_{\rm BR}$ is $\sim 24\pm1$ mag arcsec$^{-2}$ in both the late type dwarfs and the more massive spiral galaxies (their Fig. 8 and Table 5), which is compatible with our results. A lack of correlation between the surface brightness at the break radius and the galaxy stellar mass, combined with the fact that the breaks appear at about two magnitudes brighter surface brightness levels than the limit of our data (see Table \ref{table:sigmasky} and Fig. \ref{ii_iii_trad} panel f), also indicates that the breaks can be observed similarly in the low- and high-mass galaxies (similar conclusion was made by \citealt{salo2015}, their Fig. 31). This also means that the dependency of profile type with stellar mass (shown in Fig. \ref{ir_sample_types}), is not an artefact of insufficiently deep data.

The ratio between the inner and outer disc scalelength ($h_{\rm i} / h_{\rm o}$) and that of the break radius and the disc scalelength inside the break ($R_{\rm BR} / h_{\rm i}$) are often used as measures of the strength of the break. We find that using both parameters, the Type II break strength slightly increase with stellar mass, whereas the break strengths of Type III profiles are roughly constant throughout the galaxy stellar mass range (Fig. \ref{ii_iii_trad} panels g and h).

\begin{figure*}
    \begin{center}
        \includegraphics[width=\textwidth]{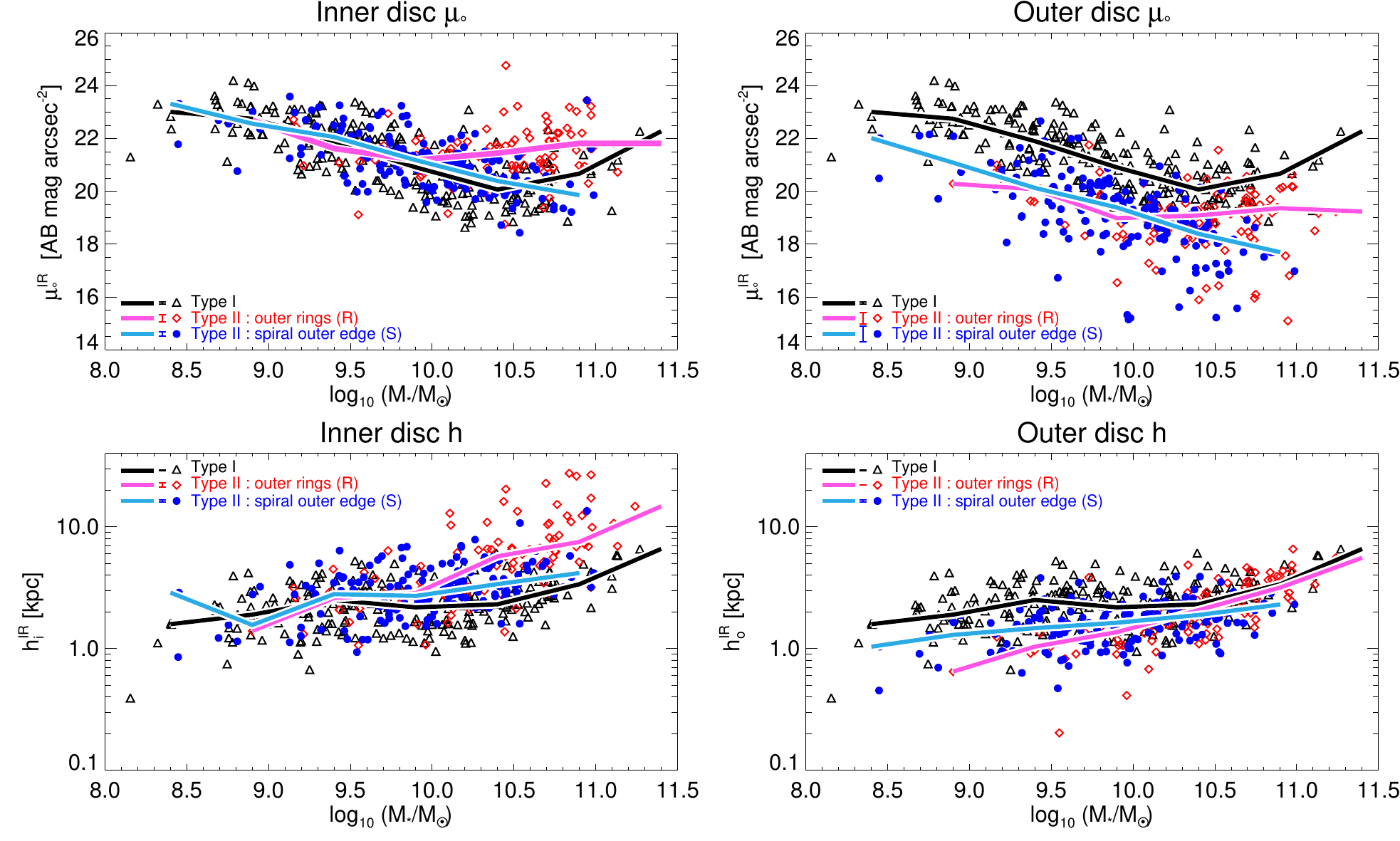}  
        \caption{As Figure \ref{ii_iii_trad} upper two rows, but now the parameters of Type II have been separated into the two main groups: Type II breaks connected with outer rings (R), and Type II breaks connected with the visual spiral outer edge (S).}
        \label{ii_param_reason}
    \end{center}
\end{figure*}

\begin{figure*}
    \begin{center}
        \includegraphics[width=\textwidth]{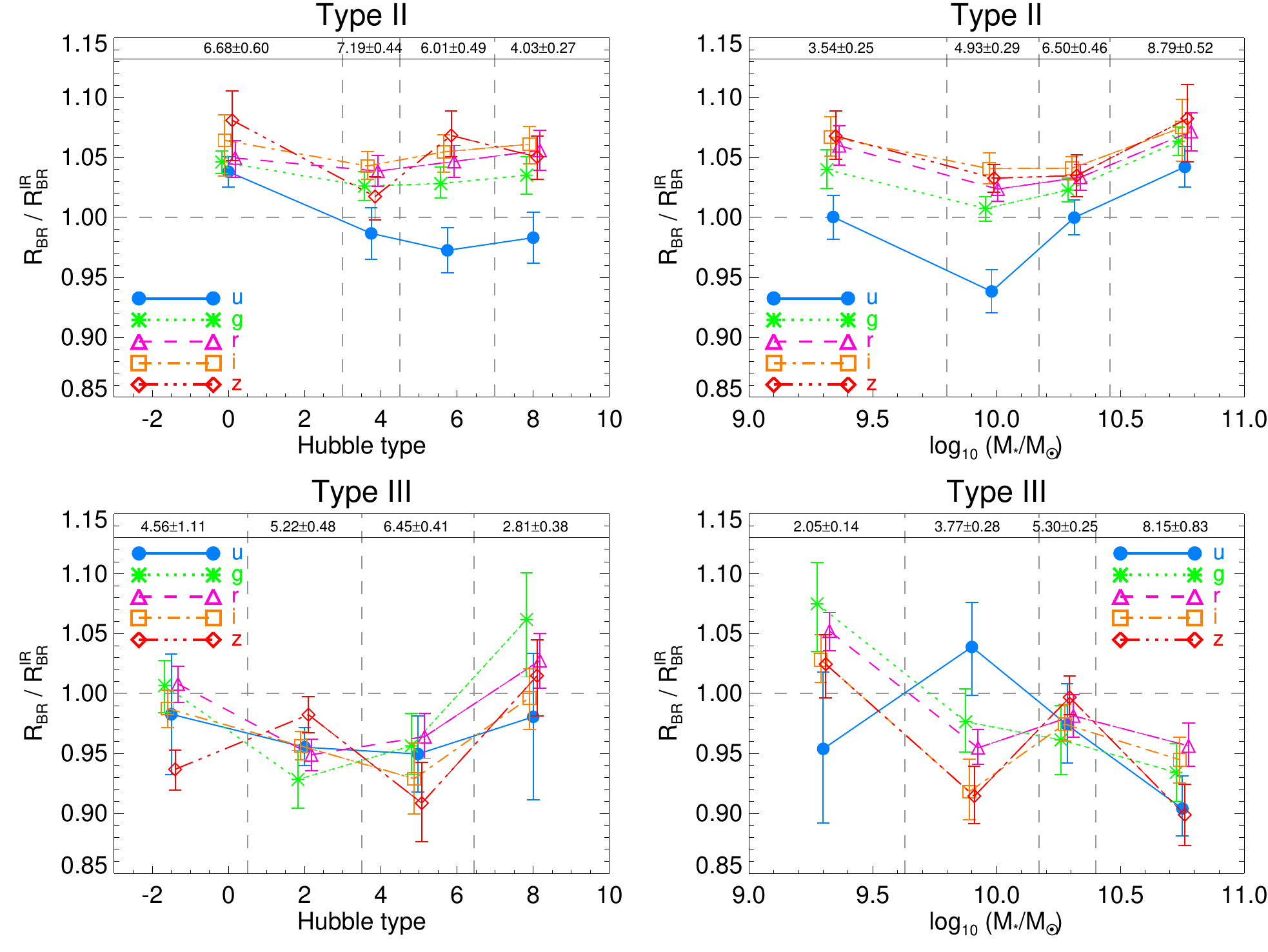}
        \caption{Break radius ($R_{\rm BR}$) in different bands, normalized with the infrared $R_{\rm BR}^{IR}$, shown separately for Type II (\emph{upper panels}) and III breaks (\emph{lower panels}). In the left panels we show the behaviour as a function of Hubble types, and in the right panels as a function of stellar masses. In every panel the galaxies are divided into four bins with roughly the same number of galaxies, and the mean value is calculated in each bin. The bin limits are shown with vertical grey dashed lines. The error bars indicate bootstrapped $\pm 1 \sigma$ uncertainty of the mean. The numerical values in the box at the top of each bin indicate the mean value of the infrared data in that bin, and  bootstrapped $\pm 1 \sigma$ uncertainty of the mean.}
        \label{trad_diff}
    \end{center}
\end{figure*}

\subsubsection{Connection with the galaxy structure components}
\label{struct_connection}

Using the measurements of rings, lenses, and spiral structures, given by \cite{laurikainen2011}, \cite{comeron2014}, and \cite{herreraendoqui2015} for our sample galaxies, we connected the break radii with these structures. As in \citeauthor{laine2014} (2014, discussed also in Section \ref{break_types}) we have divided the Type II breaks into two main groups: (a) breaks associated with rings and lenses (inner r/l or outer R/L, 2\% and 33\%  of Type II breaks, respectively), and (b) those associated either with bright star-formation regions inside the spiral arms (BS, 10\% of Type II breaks), or with visual outer edges of the spiral arms (S, 47\% of Type II breaks. Our group of outer rings and lenses (R and L) is fundamentally the same as the original Type II.o-OLR introduced by \cite{pohlen2006} and \cite{erwin2008}. The breaks located at the visual outer edges of the spirals (S) are comparable to the Type II.o-CT of \cite{pohlen2006} and \cite{erwin2008}. In 5\% of Type II profiles the break can be associated with asymmetric outer discs (AB). These are mainly found in the $T \gtrsim 7$ galaxies. Only in 3\% of Type II profiles no visual association can be made with particular galaxy structure component (NA). In Figure \ref{ii_mass_connection} (upper panels) we show how these groups behave as a function of galaxy mass, and Hubble type. As expected, the Type II breaks associated with rings and lenses are more common among the high-mass galaxies. On the other hand, in the lower mass galaxies the breaks associated with the spiral arms are more common. This mass segregation of the two main Type II break groups is consistent with their division among the Hubble types, found also in previous studies (e.g. \citealt{pohlen2006}, see also \citealt{laine2014} and Fig. \ref{ii_mass_connection} upper right panel). We do not find any systematic association of the structure components with Type III break radii.

In Figure \ref{ii_mass_connection} we show for Type II profiles how $R_{\rm BR}^{IR}$ correlates with the radius of the structure the break is connected with. The radius, both of the ring (r, R) and lens (L) structures, coincide well with  $R_{\rm BR}^{IR}$ (Fig. \ref{ii_mass_connection} lower left panel). In the lower right panel we compare $R_{\rm BR}^{IR}$ with the visual outer edge of the spirals (S), and with the radius where the spiral arms vanish in the outer disc, which radius was estimated from the logarithmic spiral section fits by \cite{herreraendoqui2015}. $R_{\rm BR}^{IR}$ connected with spiral structure outskirts (S) agrees well with the maximum radius of the spiral arms. Furthermore, the breaks associated with radial distances of the bright star formation regions inside the spiral arms (BS) appear at a radius which is approximately 0.5 times smaller than the maximum radii of the spiral arms.

In Figure \ref{ii_param_reason} we show the central surface brightnesses and scalelengths of the discs for Type II profiles in which the breaks are associated with outer rings (R), and those associated with the visual outer edges of the spiral arms (S). Among the galaxies with masses $\log_{10} (M_{*}/M_{\sun}) \lesssim 10$, the medians of $\mu_{\circ}^{IR}$ and $h^{IR}$ of the inner discs of these two groups are very similar to those of Type I. Moreover, the largest differences relative to Type I profiles for these Type II groups appear in $\mu_{\circ}^{IR}$. Both have increasing break radius with stellar mass (i.e. behaviour seen in Fig. \ref{ii_iii_trad} panel e), and the average surface brightness at the break radius is also similar ($23.4 \pm 1.1$ and $23.5 \pm 1.0$, respectively for those associated with outer rings and spiral outer edge). However, the Type II breaks connected with outer rings are on average slightly stronger ($h_{\rm i}^{IR} / h_{\rm o}^{IR} \approx 2.8 \pm 1.5$ vs. $2.0 \pm 0.8$), and have smaller values of $R_{\rm BR}^{IR} / h_{\rm i}^{IR}$ ($1.7 \pm 0.8$ vs. $2.2 \pm 0.8$), than those associated with the visual spiral outer edges.

\subsection{Parameters of the discs as a function of observed wavelength band}

We now focus on studying how the scalelength of the disc and the properties of the breaks behave as a function of wavelength band. Here we include only cases where the profile type classification is the same in all six bands, and the image depth is sufficient for a reliable determination of the disc parameters outside the breaks. We considered only the main Types I, II, and III having only one break in the profile. Mixed types with two breaks are not included here, or are the Type II.i profiles. In total 239 galaxies were examined, out of the 258 galaxies in which the profile type is the same in every band. As the infrared data is a reasonable estimate of the stellar mass distribution of the galaxy, it was used as a baseline for the comparison of the parameters in the other bands.

\subsubsection{Break radius}

In Figure \ref{trad_diff} we show how the Type II break radius ($R_{\rm BR}$) in the different bands compares with that in the infrared, among all Hubble types (upper left panel) and stellar masses (upper right panel). The breaks in the \emph{u}-band appear slightly more inside in the galaxies than in the other optical bands, especially among the galaxies with later Hubble types. On the other hand, in \emph{g}, \emph{r}, \emph{i}, and \emph{z} bands the breaks appear at similar radial distances, which is slightly larger than the radius in the infrared. These differences of $\sim 5\%$ relative to the infrared translates to a mean difference of $\sim 2$ arcseconds, which means $\approx 3$ pixels at $3.6 \, \mu m$, and $\approx 5$ pixels at optical SDSS data. However, these differences are within the measurement uncertainties of $R_{\rm BR}$, with contributing factors from different pixel sizes, the limitations of the disc profile fitting methods, and the different FWHM of the data. In the \emph{u}-band the difference is real and it is possible that light could be smeared in the longer wavelengths, due to the increased velocity dispersion and radial migration of the old stars, thus altering the break radius at longer wavelengths. The behaviour of $R_{\rm BR}$ remains similar when it is shown separately for the Type II breaks associated with the outer rings and spiral outer edges (Fig. \ref{ii_ct_olr_diff} upper left panel). However, $R_{BR}$ in \emph{u}-band for Type II, associated with the visual spiral outer edge (S), is more clearly inside the $R_{\rm BR}^{IR}$.

In all the SDSS bands the Type III $R_{\rm BR}$ is slightly more inside than that in the infrared (Fig. \ref{trad_diff} lower panels). No strong variations appear as a function of Hubble type and galaxy stellar mass. The differences in $R_{\rm BR}$ are largest for Hubble types $T \sim 0-6$, whereas no differences appear for $T<0$ or $T>6$.

\begin{figure*}
    \begin{center}
        \includegraphics[width=\textwidth]{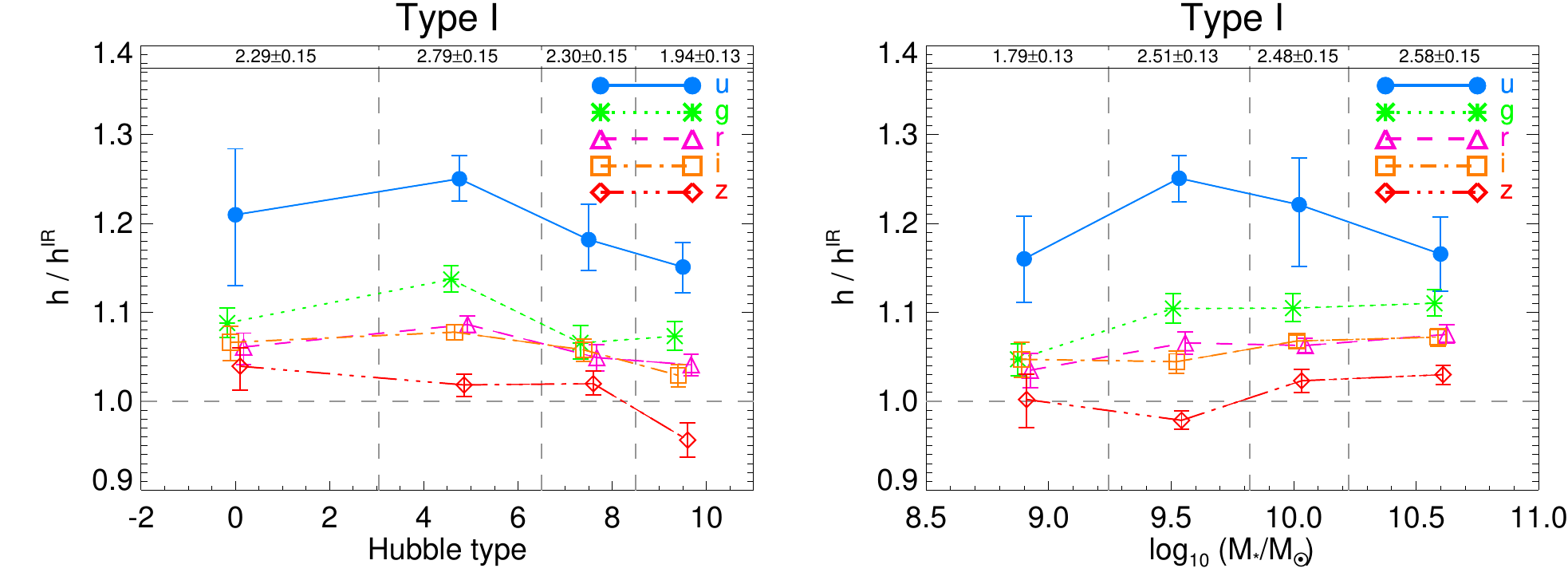}
        \caption{Single exponential Type I profile scalelength $h$, normalized to $h$ at infrared ($h/h^{IR}$), plotted as a function of Hubble type (\emph{left panel}) and as a function of stellar mass (\emph{right panel}). Definition of the error bars etc. are given in the Figure \ref{trad_diff} caption.}
        \label{i_scale_diff}
    \end{center}
\end{figure*}

\subsubsection{Disc scalelengths}

\subsubsection*{Type I}

In Figure \ref{i_scale_diff} we compare the disc scalelengths of Type I profiles in different wavelength bands, normalized to the scalelength in the infrared ($h / h^{IR}$). We see a systematic behaviour of increasing $h / h^{IR}$ with decreasing wavelength. There is also a morphological dependency: in Hubble types $T \gtrsim 6$ the ratio $h / h^{IR}$ decreases. It is worth noticing that the lowest mass galaxies (generally late-types) have larger gas-to-stellar mass ratios and higher specific star formation rates than the more massive galaxies (e.g. \citealt{brown2015}). This star formation comes through even at $3.6 \, \mu m$, which has a contribution from polycyclic aromatic hydrocarbons and hot dust emission from star formation regions \citep{meidt2012}, which could explain the decrease in $h / h^{IR}$ in the latest Hubble types. In types $T<6$ $h / h^{IR}$ is nearly constant in each band. As a function of stellar mass $h/ h^{IR}$ is nearly constant in \emph{u}-band (Fig. \ref{i_scale_diff} right panel), while in the other bands it slightly decreases towards lower stellar masses. 

Earlier studies also reported that $h$ depends on the observed wavelength band (e.g \citealt{dejong1996a,macarthur2003,fathi2010}). \cite{fathi2010} found, in their sample of 30 000 spiral galaxies ($1 \leq T \leq 8$), that the average $h$ has an increase of $35 \%$ from \emph{r} to \emph{u}-band. This is clearly a larger increase than the $18\pm4$\% that we find, which could be because we include only discs with Type I profiles here.

\subsubsection*{Type II}

\begin{figure*}
  \begin{center}
    \includegraphics[width=\textwidth]{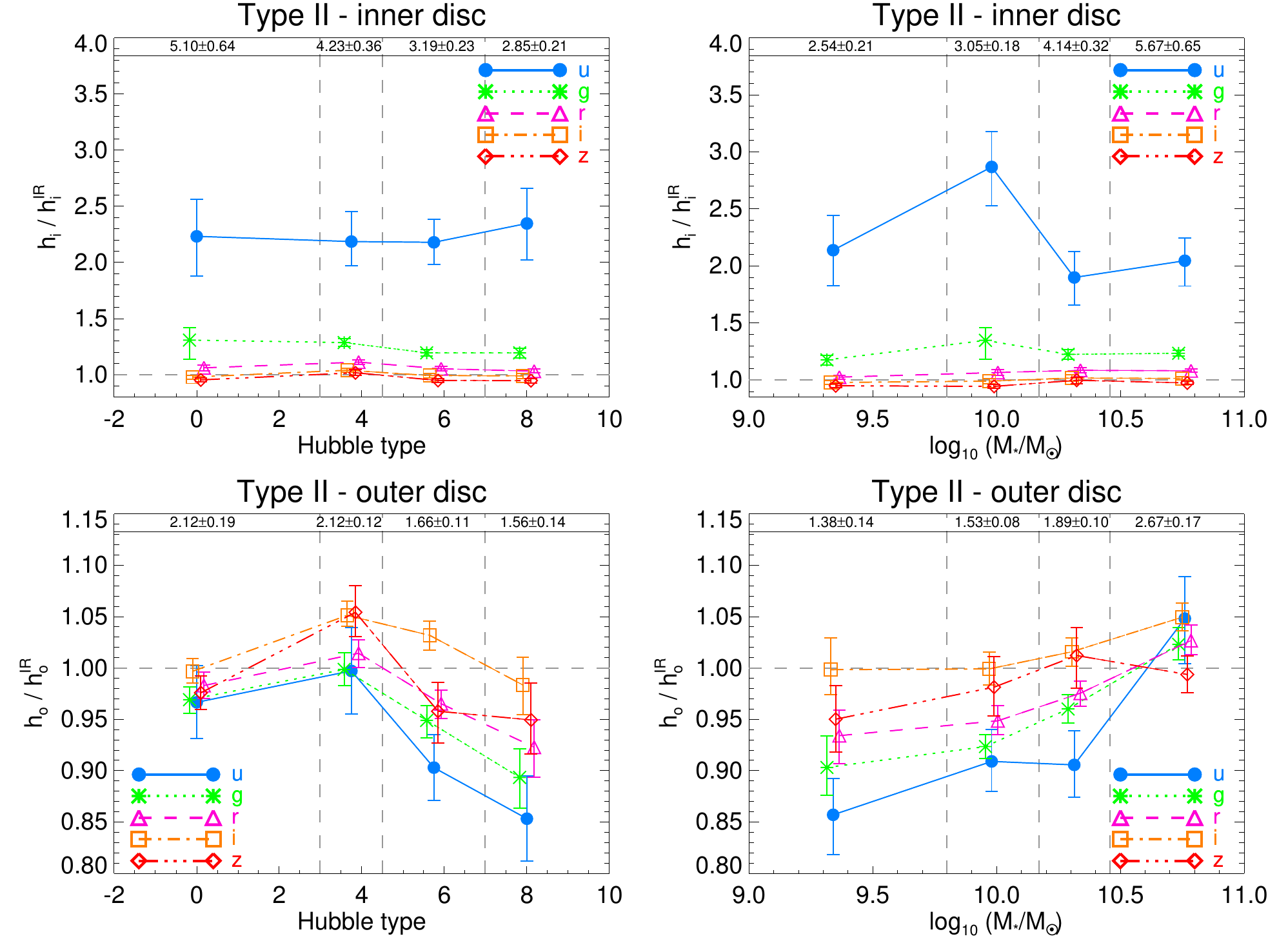}
    \caption{As Figure \ref{i_scale_diff}, but the scalelengths of the discs inside ($h_{\rm i}$, \emph{upper panels}) and outside ($h_{\rm o}$, \emph{lower panels}) of Type II breaks are shown separately.}
    \label{ii_scale_diff}
  \end{center}
\end{figure*}

In Figure \ref{ii_scale_diff} we show the scalelengths of the discs for Type II breaks, as a function of Hubble type and stellar mass. Shown separately are the relative scalelengths for the inner ($h_{\rm i} / h_{\rm i}^{IR}$, upper panels) and outer ($h_{\rm o} / h_{\rm o}^{IR}$, lower panels) discs. Most importantly we find that  $h_{\rm i}^u$ is on average $\sim 2.2$ times larger than $h_{\rm i}^{IR}$, throughout the Hubble type and stellar mass ranges. Also, $h_{\rm i}^g$ is a factor of $\sim 1.3$ larger than $h_{\rm i}^{IR}$. In \emph{r}, \emph{i}, and \emph{z} bands the parameter $h_{\rm i}$ is the same as in the infrared.

There is also a weak dependency between $h_{\rm o} / h_{\rm o}^{IR}$ and the observed wavelength band (Fig. \ref{ii_scale_diff} lower panels): the ratio decreases in Hubble types $T \gtrsim 5$ (lower left panel) and at lower stellar masses ($\log_{10} (M_* / M_{\sun}) \lesssim 10.5$, lower right panel) especially in \emph{g} and \emph{u} bands. In the two latest Hubble type bins, and in the two lowest stellar mass bins, with increasing wavelength $h_{\rm o}$ approaches the values of $h_{\rm o}^{IR}$. The change of $h_{\rm i}$ and $h_{\rm o}$  as a function of wavelength in the late-type and low-mass galaxies is in qualitative agreement with the average colour profiles studied by \citet{bakos2008}. For Type II galaxies they found U-shaped colour profiles which show the bluest colours at the break radius, followed by reddening with increasing radius.

\begin{figure*}
    \begin{center}
        \includegraphics[width=\textwidth]{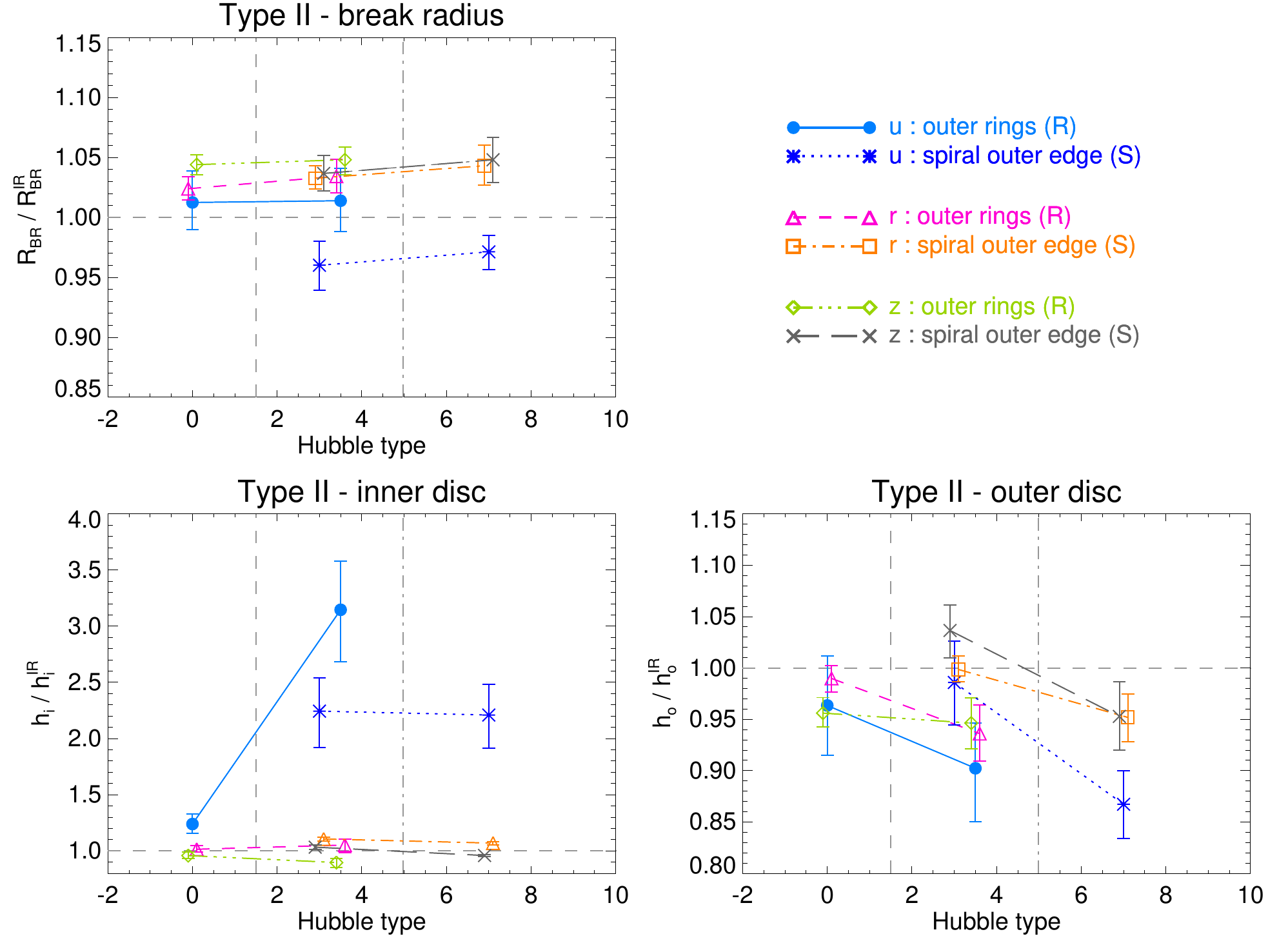}
        \caption{Mean values of $R_{\rm BR} / R_{\rm BR}^{IR}$, $h_{\rm i} / h_{\rm i}^{IR}$, and $h_{\rm o} / h_{\rm o}^{IR}$ for \emph{u}, \emph{r}, and \emph{z} bands separately for the two Type II groups: breaks connected with outer rings, and breaks connected with the visual spiral outer edge. Both groups are separated into two bins by their median Hubble type, with the bin dividing values shown as vertical dashed and dot-dashed lines. The error bars indicate bootstrapped $\pm 1 \sigma$ uncertainties of the mean.}
        \label{ii_ct_olr_diff}
    \end{center}
\end{figure*}

In Figure \ref{ii_ct_olr_diff} (lower panels) we show $h_{\rm i} / h_{\rm i}^{IR}$ and $h_{\rm o} / h_{\rm o}^{IR}$ for \emph{u}, \emph{r}, and \emph{z} bands, separately for the two main Type II groups. The same general behaviour as seen in Figure \ref{ii_scale_diff} holds, but some differences also appear. Namely, in breaks associated with outer rings the ratio $h_{\rm i}^u / h_{\rm i}^{IR}$ increases strongly from $T\lesssim1.5$ to $T\gtrsim1.5$. This indicates a contribution of star formation in the outer rings of spiral galaxies, in which galaxies $h_{\rm i}^u$ can be very large. In addition, the $h_{\rm o}^u$ associated with the spiral outer edge decreases relative to $h_{\rm o}^{IR}$ from $T<5$ to $T>5$ galaxies, which could indicate scarcity of young stars in the outer discs, particularly for galaxies with $T>5$.

\subsubsection*{Type III}

\begin{figure*}
  \begin{center}
    \includegraphics[width=\textwidth]{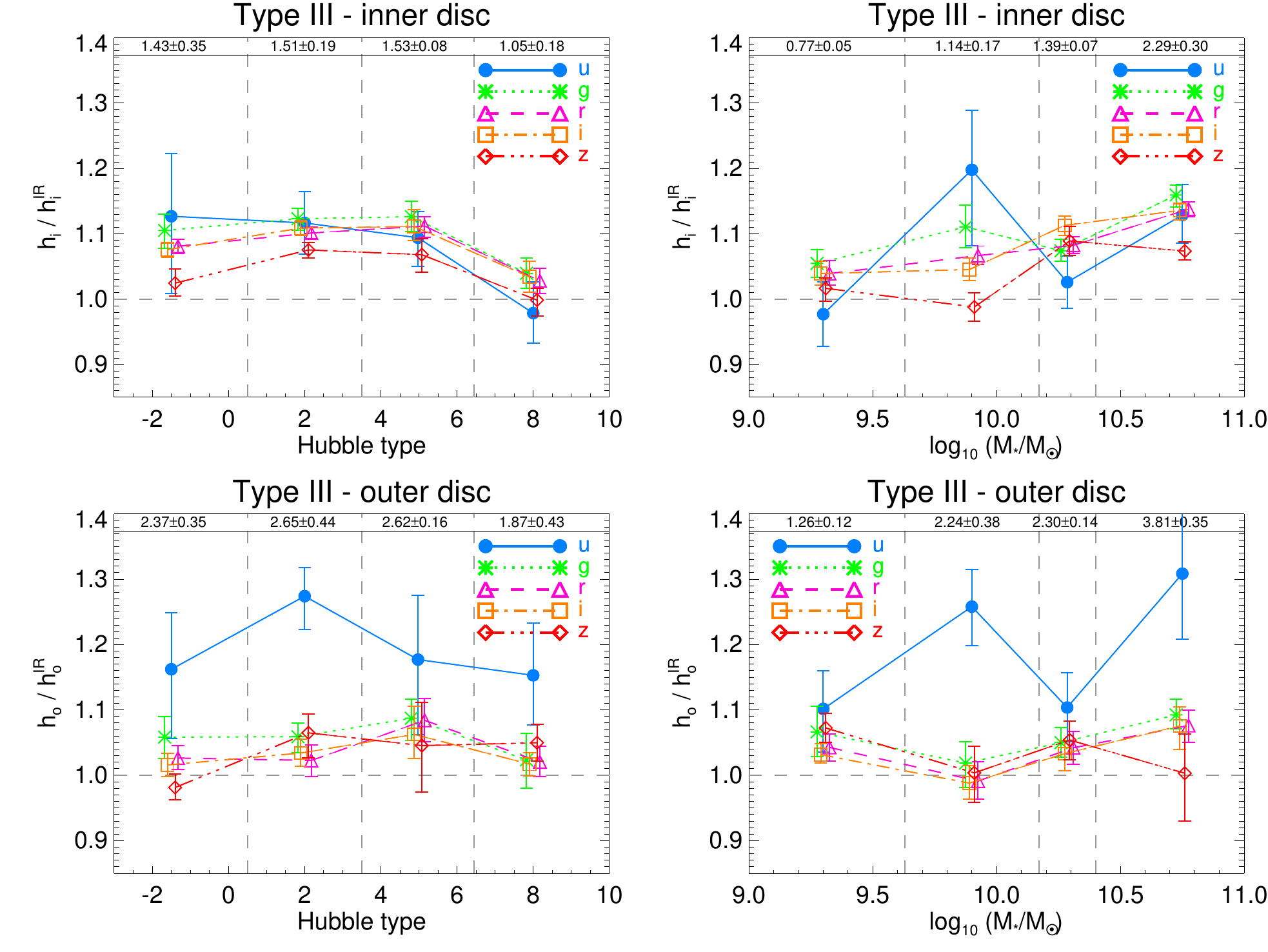}
    \caption{As Figure \ref{i_scale_diff}, but the scalelengths of the discs inside ($h_{\rm i}$, \emph{upper panels}) and outside ($h_{\rm o}$, \emph{lower panels}) of Type III breaks are shown separately.}
    \label{iii_scale_diff}
  \end{center}
\end{figure*}

We show similar plots for Type III profiles in Figure \ref{iii_scale_diff} as we did for Type II profiles in Figure \ref{ii_scale_diff}. It appears that in massive galaxies $h_{\rm i}$ is larger in the optical than in the infrared (upper right panel), but in low mass galaxies $h_{\rm i}$ is the same in all bands. However, these changes in $h_{\rm i}$ of Type III profiles between the different bands are small compared to the changes in $h_{\rm i}$ in Type II profiles.

In the outer discs of Type III profiles $h_{\rm o}^u$ is larger than in the other bands in all Hubble types (Fig. \ref{iii_scale_diff} lower panels). In all the other optical bands $h_{\rm o} \approx h_{\rm o}^{IR}$.

\subsubsection{Break strength}

The fact that $h_{\rm i}$ and $h_{\rm o}$ of Type II and III breaks depend on wavelength band means that $h_{\rm i} / h_{\rm o}$, used as a measure of the break strength, also depends on wavelength. Namely, for Type II profiles in \emph{u} band  $h_{\rm i} / h_{\rm o}$ is on average a factor of $\sim 2.5$ larger than in infrared. The break strength then quickly decreases so that in \emph{g} band $h_{\rm i} / h_{\rm o}$ is larger only by a factor $\sim 1.3$ compared to infrared. In \emph{i} and \emph{z} bands $h_{\rm i} / h_{\rm o}$ is the same as in infrared. For Type II profiles there is also a small morphological dependency so that in \emph{u} band in late-type ($T \gtrsim 5$) and low-mass galaxies ($\log_{10} (M_* / M_{\sun}) \lesssim 10.5$) the breaks are stronger than in the more massive galaxies in earlier Hubble types. This is reflected also in the behaviour of the disc scalelengths in Figure \ref{ii_scale_diff}.

For Type III profiles there is no strong change in the break strength among the different bands.  In \emph{z}, \emph{i}, \emph{r}, and \emph{g} bands the Type III breaks are slightly less strong than in the infrared ($h_{\rm i} /h_{\rm o}$ a factor $\sim 1.05$ different), while in the \emph{u} band the Type III breaks are slightly stronger ($h_{\rm i} /h_{\rm o}$ a factor $\sim 0.95$ different).

Dependency of $h_{\rm i} / h_{\rm o}$ on the observed wavelength band has been reported previously for Type II breaks by \citet{bakos2012}. They study four Sb-Sc galaxies, using methods comparable to our study, to derive the disc parameters. In addition, \citet{radburnsmith2012} show, using star count measurements of the Sc galaxy NGC 7793, that the Type II break in this galaxy is smoother for older stellar populations. We have shown that Type II break strength is more dependent on the large changes in $h_{\rm i}$, than on those in $h_{\rm o}$.


\section{Discussion}
\label{discussion}

\subsection{Galaxy stellar mass constraining Type II break formation}
\label{discuss:mass}

We have shown that the fractions of Type I and II breaks depend on the stellar masses of the galaxies: the fraction of Type I profiles increases, and that of Type II decreases with decreasing stellar mass (see Fig. \ref{ir_sample_types} right panel). This is also reflected in the distribution of the profile types among the Hubble types: the single exponential Type I profiles are particularly common in late-types ($T \gtrsim 6$, Fig. \ref{ir_sample_types} left panel). As stellar mass is perhaps the most fundamental parameter of a galaxy, this raises an important question: are there processes in lower mass galaxies that inhibit the break formation, or alternatively, are the breaks forming in galaxies as they grow in mass? This is intriguing especially when the origin of exponential discs is not yet fully understood (see for example the review by \citealt{vanderkruit2011}). 

\begin{figure*}
  \begin{center}
    \includegraphics[width=\textwidth]{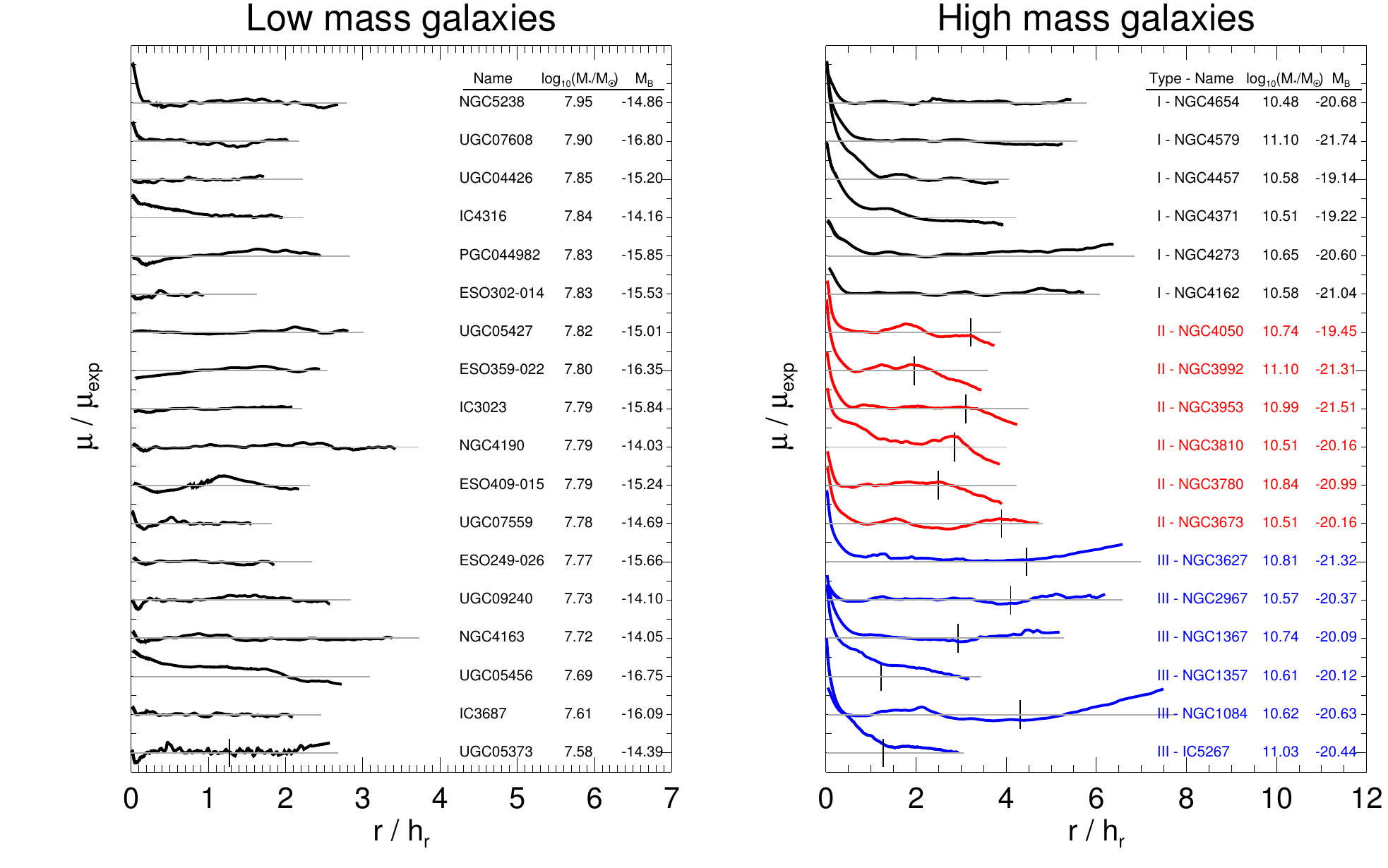}
    \caption{$3.6 \, \mu m$ radial surface brightness profiles normalized with the exponential disc surface brightness ($\rm \mu_{exp}$), and disc scalelengths ($\rm h_r$), of the two dimensional decomposition values from Spitzer Survey of Stellar Structure in Galaxies Pipeline 4 (S$^4$G P4, \citealt{salo2015}). A single exponential disc would appear as horizontal line in this plot (grey lines). In \emph{the left panel} we show a random sample of 18 low mass galaxies, which are not part of this study but fulfil the inclination criteria of the sample selection ($b/a > 0.5$). In \emph{the right panel} we show random sample of 18 high mass galaxies with Type I, II, and III disc profiles, which are part of this study. In the high mass galaxies we show the location of the Type II or III break by a short vertical line segment. All the profiles have been cut by the $\mu = 25.5$ mag arcsec$^{-2}$ radius of the $3.6 \, \mu m$ data.}
    \label{low_hi_prof}
  \end{center}
\end{figure*}

The surface brightness in the infrared where the Type II breaks appear, is uncorrelated with the galaxy stellar mass in our sample (see Fig. \ref{ii_iii_trad} panel f). That is the case also with the strength of the break (i.e. $h_{\rm i} /h_{\rm o}$, see Fig. \ref{ii_iii_trad} panel g). Since $\mu_{\rm BR}^{IR}$ is roughly two magnitudes brighter than the depth of our data (see Fig. \ref{ii_iii_trad} panel f, and Table \ref{table:sigmasky}) indicates that we are not misclassifying Type II profiles as Type I in the low-mass galaxies, and also that the observed trends with mass are not imprints of smoother breaks or of breaks appearing at lower surface brightness levels. The trend of Type I becoming more prevalent in lower masses is further supported by Figure \ref{low_hi_prof}. In this figure we show a random collection of normalized surface brightness profiles of low-mass S$^4$G galaxies (left panel) which are not included in our sample (i.e. they have $T > 9$), but still fulfil our sample selection criterion $b/a > 0.5$. The used normalization does not change the location or strength of the breaks, but makes it easier to compare with a single exponential disc, which would appear as a nearly horizontal lines in this figure. For comparison, in the right panel of Figure \ref{low_hi_prof} we show  high-mass galaxies which are part of this study. Detailed disc profile analysis has not been done on these extra low mass galaxies ($-17 < M_{\rm B} < -14$, roughly equal to $7.5 \lesssim \log_{10} ( M_* / M_{\sun}) \lesssim 8.0$), but they do possess on average a single exponential surface brightness profile. Some exceptions are naturally found, for example ESO 409-015 and IC 4316. On the other hand, \cite{herrmann2013} have demonstrated, using mostly \emph{U}-, \emph{B}-, and \emph{V}-band data, that even $67 \%$ of dwarf irregular galaxies have a Type II break (see also \citealt{hunter2006,hunter2011}). Their dwarf galaxies have $M_{\rm B}$ in the same range as the low mass galaxies shown here in Figure \ref{low_hi_prof}. Dwarf irregular galaxies have a morphology dominated by star formation clumps, which could manifest as breaks in the surface brightness profiles at shorter wavelengths. Thus, in dwarf irregular galaxies the Type II breaks could be more common at optical bands, even if in infrared their discs have Type I profiles. However, further studies are needed to fully explain the discrepancy between our results and those of \cite{herrmann2013}.

\subsubsection{Bar induced Type II break formation}

In the high-mass galaxies ($\log_{10} (M_{*}/M_{\sun})>10.5$), and early Hubble types ($T \lesssim 2$), Type II breaks are mostly associated with outer rings (53\% and 74\% respectively, 30 \% of all Type II in our sample, see Fig. \ref{ii_mass_connection}, and Fig. 7 in \citealt{laine2014}). The outer rings are thought to form through gas accumulation at the Outer Lindblad Resonance of the bar, which has resulted to enhanced star formation and the emergence of stellar outer ring (e.g. \citealt{buta1996}). In the early types ($T \lesssim 2$) and generally massive galaxies, outer rings are commonly found \citep{buta1996,comeron2014,herreraendoqui2015}. The reduced frequency of Type II breaks associated with these structures in low mass galaxies (i.e. $\log_{10} (M_{*}/M_{\sun})<10.5$, see Fig. \ref{ii_mass_connection}) could explain the higher fraction of Type I profiles in low mass galaxies (together with the decreasing fraction of Type III profiles towards lower masses).

In the high-mass and early-type galaxies the formation of Type II breaks would then result from the angular momentum redistribution caused by the bar, followed by subsequent outer ring formation (e.g. \citealt{pohlen2006,debattista2006,foyle2008,minchev2012}). This is consistent with our finding that, compared to infrared, the \emph{u}-band scalelengths $h_{\rm i}^{u}$ rise by a factor of $\sim$ 1.3 in galaxies with Hubble types $T < 2$. In Hubble types $T>2$ star formation in the outer rings seems to change the properties of the inner disc more strongly, so that $h_{\rm i}^u$ is even by a factor of $\sim$ 2.9 larger than $h_{\rm i}^{IR}$ (Fig. \ref{ii_ct_olr_diff} lower left panel). On the other hand, the scalelengths of the outer discs ($h_{\rm o}$) do not depend on the observed wavelength band (Fig. \ref{ii_ct_olr_diff} lower right panel). In bright early-type galaxies the behaviour of $h_{\rm i}$ could be interpreted to arise from star formation in the outer rings, causing an accumulation of young stars in the inner disc. Star-forming outer rings have been indeed observed in bright early-type galaxies (e.g. \citealt{salim2012}).

\cite{foyle2008} predict, based on their N-body simulations, that if the break is associated with bar-induced dynamical effects, the inner discs evolve with the bar, whereas the outer density profiles (i.e. $h_{\rm o}$) would resemble the initial discs before the break formation. Our observations are consistent with that idea: we have shown that $h_{\rm o}$ of Type II profiles with outer rings (which are bar related features) are similar with the scalelengths of Type I profiles. This is the case for the bright galaxies with $\log_{10} (M_{*}/M_{\sun}) \gtrsim 10.5$, although the models of \cite{foyle2008} do not match our observed properties of low mass galaxies. Namely, we find that in the lower mass galaxies ($\log_{10} (M_{*}/M_{\sun}) \lesssim 10$) the scalelengths of Type I profiles are larger than $h_{\rm o}^{IR}$ of Type II breaks. Moreover, in the models by \cite{foyle2008} a critical parameter for the break formation is $m_d$/$\lambda$, where $m_d$ is disc-to-total mass ratio and $\lambda$ is halo spin parameter: the disc is stable with small $m_d$/$\lambda$ values, in which case fairly exponential disks are expected. Galaxies with small $m_d$/$\lambda$ corresponds to the low mass and lower surface brightness galaxies in our sample, which galaxies indeed have more frequently single exponential surface brightness profiles. However, it is worth noticing that Type II profiles also appear in some low mass galaxies. 

Curiously, in a few high-mass galaxies, the Type II break is associated with an outer lens ($3\%$ of all Type II breaks), or with an inner ring in an unbarred galaxy ($2\%$ of Type II breaks). These breaks are rare and do not significantly alter the average behaviour of the disc scalelengths discussed above. In fact, the outer lenses are also suggested to be assigned to bar resonances (e.g. \citealt{laurikainen2013}). Usually lenses show no star formation, and in such cases the scalelengths of the discs do not change dramatically with wavelength band. However, it is unclear how the inner rings in unbarred galaxies were formed and many possible scenarios have been proposed, including minor merging (e.g. \citealt{elichemoral2011}). Breaks associated with these inner rings show similar increased $h_{\rm i}$ in the \emph{u} band as the breaks associated with the outer rings. The immediate explanation might be the same, namely star formation.

\subsubsection{Type II formation through star formation threshold and radial migration}

In low-mass galaxies ($\log_{10} (M_{*}/M_{\sun}) < 10.5$) the Type II breaks are more often associated with the visual outer edges of spiral arms (55\% of Type II breaks in galaxies with $\log_{10} (M_{*}/M_{\sun}) < 10.5$, 47\% of all Type II in our sample, see Fig. \ref{ii_mass_connection}). These Type II breaks are largely found in late-type galaxies ($T \gtrsim 2$, \citealt{pohlen2006, gutierrez2011, laine2014}, see also Fig. \ref{ii_mass_connection}). The breaks in late-type galaxies are generally thought to have formed by a star-formation threshold, beyond which star formation is reduced (e.g. \citealt{schaye2004,elmegreen2006,pohlen2006} and references therein). The location of the star-formation threshold in the disc depends on the surface mass density of the cold gas \citep{schaye2004,roskar2008a}, or alternatively on the gas volume density as a result of a warp in the gas disc \citep{sanchezblazquez2009}. \cite{azzollini2008b} have revealed that in $z \sim 1$ galaxies the breaks are located much closer to the centre than in the local Universe. They argue that the evolution in the break position could be explained by the star formation threshold moving to larger radii as the galaxies grow in an inside-out fashion.

In the late-type spiral galaxies with Hubble types $T \gtrsim 2$ radial migration could further modify the outer disc. The migration could be induced by the transient spiral structure (e.g. \citealt{sellwood2002,roskar2008a,martinezserrano2009,roskar2012}), bar-spiral resonance coupling (e.g. \citealt{minchev2010}), or satellite galaxy interactions (e.g. \citealt{quillen2009,bird2012}). Migration is a slow random process, and the consequences should be best observed when comparing the youngest ($\sim$ \emph{u}-band) and the oldest stellar populations ($\sim$ infrared). 

Our results show that in galaxies having  $\log_{10} (M_{*}/M_{\sun}) < 10.5$ and $4 \lesssim T \lesssim 9$, $h_{\rm i}$ is a factor of $\sim 2.2$ larger in the \emph{u}-band than in the infrared, indicating the presence of young stellar populations in the inner discs (Fig. \ref{ii_scale_diff} upper panels). In addition, in these galaxies there is a weak systematic behaviour in $h_{\rm o}$: it is slightly shorter in bluer wavelengths, and the difference to infrared increases when moving to later Hubble types and lower stellar masses (i.e. in $T \approx 8$, $h_{\rm o}$ is on average 10\% smaller in \emph{u}-band compared to infrared, Fig. \ref{ii_scale_diff} lower panels). The same behaviour in $h_{\rm i}$ and $h_{\rm o}$ appears when studying separately the Type II breaks associated with the visual spiral outer edges (see Fig. \ref{ii_ct_olr_diff}). This behaviour of the disc scalelengths in late-type $T \gtrsim 4$ galaxies is in qualitative agreement with the star-formation threshold creating a break in the surface brightness profile, with radial migration modifying the properties of the outer disc. The radial migration could populate the outer discs with older stars, which have had more time for the excursion, therefore increasing $h_{\rm o}$ at longer wavelengths. The radial migration could also be working more effectively in the latest galaxy types and lowest stellar masses ($T \approx 8$, $\log_{10} (M_{*}/M_{\sun}) \approx 9.5$), where the largest changes in $h_{\rm o}$ are seen between the different bands. However, the effect of galaxy morphology on the radial migration process has not received much attention in simulations.

Star count observations of the Sc galaxy, NGC 7793, by \cite{radburnsmith2012} have similarly shown that the scalelength of the inner disc increases, while that of the outer disc decreases, towards younger stellar populations. This is comparable with the behaviour described above for the late-type galaxies with Type II breaks in this study. \cite{radburnsmith2012} further compare their observations with N-body simulations and argue that stellar radial migration indeed is responsible for the increasing scalelength of the outer disc.  Observations of \citet{marino2015} have also revealed that Type II profiles in low-mass galaxies ($< 10^{10} M_{\sun}$) are associated with weaker metallicity gradients in the disc, compared to Type II profiles in more massive galaxies. They suggested that stellar radial migration is causing this metallicity flattening in the low-mass galaxies. The increase in $h_{\rm o}$ at longer wavelengths is also comparable with the U-shaped colour and stellar age profiles observed for some of the Type II galaxies, which show bluest colour and age minima at the break radius, as well as disc reddening and increasing stellar ages outside the break \citep{azzollini2008a,bakos2008,yoachim2010,yoachim2012,roediger2012,zheng2015}. 

In principle it is possible that the star-formation threshold plays a smaller role for creating the disc breaks in lower stellar mass spirals, due to an increased efficiency of feedback processes that redistribute the gas. Combined with radial stellar migration weakening the break, this process could erase or even inhibit formation of Type II breaks in the infrared in these low-mass spirals, where we actually see more Type I profiles. Indeed, in these low-mass galaxies the inner discs ($h_i^{IR}$ and $\mu_{\circ \rm ,i}^{IR}$) of the Type II breaks, associated with the spiral outer edges, are similar to those of Type I profiles (i.e. $\log_{10} (M_{*}/M_{\sun}) \lesssim 10$, see Fig. \ref{ii_param_reason} left panels).

\subsection{Implications on Type III break formation}

The formation processes of Type III breaks are perhaps the least understood, and in general they are thought to have formed by environmental processes. Both minor \citep{laurikainen2001,penarrubia2006,younger2007} and major merging processes \citep{borlaff2014} have been shown to produce up-bending profiles in simulations. Indeed, some of the galaxies with Type III profiles are easily recognized to be interacting systems (e.g. NGC 474, NGC 722). Also, \cite{laine2014} have previously found that $h_{\rm i}$ and $h_{\rm o}$ of Type III profiles of high mass galaxies correlate with the Dahari-parameter, estimating the tidal interaction strength imposed by the neighbouring galaxies. With higher tidal interaction strength the scalelengths increase, which could result from environmental effects modifying the disc profile. On the other hand, in the simulations of \citet{minchev2012} smooth gas-accretion destabilizes the outer disc, thus spreading the disc and forming a Type III profile. Gas accretion and galaxy mergers might both contribute to the results of \cite{head2015}. They find that the Coma cluster S0s with a Type III break show an increased bulge size and luminosity compared to S0 galaxies with Type I or II profiles, which hints to a process that simultaneously causes build-up of the central mass concentration and spreading of the outer disc. Suitable star-formation conditions could also produce an up-bending profile without additional environmental effects, as shown in the semi-analytical models of \cite{elmegreen2006}. 
    
The previous studies (e.g. \citealt{pohlen2006,gutierrez2011}), including this study, have shown that the Type III profiles are more common in the early-type galaxies ($T \lesssim 3$, see Fig. \ref{ir_sample_types_highlow}). In principle the increased fraction of Type III profiles among the earlier Hubble types could be a manifestation of the morphology-density relation \citep{oemler1974,davis1976,dressler1980}, which could be further supported by the slightly increasing fraction of Type III profiles with galaxy stellar mass, as expected if mergers or gas accretion are playing an important role in creating these profiles.

What would be the observed signatures of the above processes creating the Type III profiles? Generally minor and major merging events, in which at least one of the progenitor galaxies has a supply of cold gas, boosts star formation in the remnant galaxy. Similarly, gas accretion into the discs for example via cold streams, may induce star formation and a growth of the disc (e.g. \citealt{lemonias2011,moran2012}, and references therein). Both merging and gas accretion could form an extended UV-disc (XUV-disc), which can show star formation much further out in the galaxy than the optical disc. In fact, the total fraction of Type III profiles ($24 \%$) closely matches the observed fraction of XUV-discs ($\sim 25 \%$, e.g. \citealt{gildepaz2005,thilker2005,zaritsky2007}). We performed a cross-match between our sample and the XUV-disc catalogue of \cite{thilker2007}, which revealed that 26 of our sample galaxies have XUV-discs reported in this catalogue. It appears that our disc profile types have no connection with the presence of a XUV-disc. Among these 26 galaxies the profile type fractions resemble those of our full sample (i.e. $\sim 20 \%$ Type I, $\sim 50 \%$ Type II, and $\sim 30 \%$ Type III, see also Table \ref{table:infra-stat}).

We find that Type III outer disc scalelengths ($h_{\rm o}$) are on average a factor of $\sim 1.2$ larger in the \emph{u}-band than in infrared (Fig. \ref{iii_scale_diff}). The scalelengths of the inner disc ($h_{\rm i}$) are also slightly larger at shorter wavelengths (i.e. $h_{\rm i} / h_{\rm i}^{IR} \approx 1.1$ in \emph{u}-band at $T \lesssim 6$, Fig. \ref{iii_scale_diff}). In fact, the behaviour of $h_{\rm i}$ at different bands resembles that of single exponential discs (Fig. \ref{i_scale_diff}). As such our results for the Type III profiles are compatible with a mechanism that boosts star formation, especially in the outer discs, and possibly causes an up-bend in the radial surface brightness profile. The larger scalelengths of the outer discs in the \emph{u}-band relative to the infrared, does not support the idea that the Type III up-bends in our sample were due to the stellar halo light \citep{bakos2012,martinnavarro2014,watkins2016}, or due to a superposition of thin and thick discs \citep{comeron2012}. Namely, the surface brightness profile upturns caused by stellar halos are generally found at the surface brightnesses of $\mu_{\rm r} \sim 26 - 28$ mag arcsec$^{-2}$, which is deeper than what our data can reliably probe. The thick disc and halo stars are also very old and metal poor. This means that in both cases the outer disc scalelength ($h_{\rm o}$) of the oldest populations should be considerably higher than the scalelength of the youngest populations. Our results in Fig. \ref{iii_scale_diff} do not show this kind of behaviour.

It appears that, similarly to Type II breaks, in Type III breaks the surface brightness at the break radius ($\mu_{\rm BR}^{IR}$) and the break strength ($h_{\rm i}^{IR}/h_{\rm o}^{IR}$)  are also uncorrelated with stellar mass throughout the mass range of our sample (see Fig. \ref{ii_iii_trad} panels f and g). If the Type III breaks were formed via environmental processes this scalability should constrain the properties of the interaction. The mass ratios, orbital geometries, and gas fractions of the interacting galaxies could all play an important role in explaining where in the galaxy the break is formed, or whether a break is formed at all (e.g. \citealt{younger2007}). The mechanisms responsible for such a scalability of the Type III break properties is not clear, and it is also uncertain whether the same processes are creating the up-bending light profiles in low and high-mass galaxies.


\section{Summary and conclusions}
\label{sum-conclusion}

We have studied the surface brightness profiles of 753 galaxies in a large range of Hubble types ($-2 \le T \le 9$), and stellar masses ($8.5 \lesssim \log_{10} (M_{*}/M_{\sun}), \lesssim 11$), and covering a large range of wavelength bands. We have used the $3.6 \, \mu m$ data from the Spitzer Survey of Stellar Structure in Galaxies (S$^4$G, \citealt{sheth2010}) and $K_{\rm s}$-band data from the Near Infrared S0-Sa galaxy Survey (NIRS0S, \citealt{laurikainen2011}). In addition, optical Sloan Digital Sky Survey images in five bands (\emph{u, g, r, i,} and \emph{z} bands) are used for 450 galaxies, and  Liverpool telescope \emph{g} band images for 30 galaxies. We have correlated the average behaviour of the different parameters of the discs, such as break type fractions and scalelengths, with the Hubble types and stellar masses of the galaxies. In addition, we have compared the average behaviour between the different wavelength bands.

The key results of our study are:
\begin{enumerate}
	\item The stellar mass of a galaxy affects the fractions of the different profile types: with decreasing galaxy mass the fraction of single exponential Type I profiles increases, while the fractions of Type II and Type III profiles decrease.
        \vspace{5pt}
	\item The behaviour of the fractions of the different profile types can be associated with two factors. Firstly, the frequency of the Type II breaks connected with the bar resonance structures, such as outer rings and lenses, increases with galaxy stellar mass (see Fig. \ref{ii_mass_connection}). Secondly, the Type III profiles are found more commonly among the early-type galaxies ($T \lesssim 3$, see Fig. \ref{ir_sample_types}), which are generally also more massive.
        \vspace{5pt}
	\item The observed wavelength band is an important factor in the determination of the profile type in the low mass irregular dwarf galaxies, where more Type II breaks are seen in optical (e.g. \emph{U}, \emph{B}, and \emph{V} bands) by \cite{herrmann2013}, than in the infrared in this study.
        \vspace{5pt}
	\item  The average scalelengths of the inner discs ($h_{\rm i}$) of Type II breaks are strongly affected by the observed wavelength band: they are a factor of $\sim 2.2$ larger in \emph{u} band compared to infrared. This indicates that star formation and younger stellar populations strongly affect the properties of the inner discs. The relation also holds when separating the Type II breaks into the two main subgroups: breaks connected with outer rings (R), and those connected with spiral outer edges (S). For Type II breaks connected with outer rings the increase from infrared to \emph{u} band is only a factor of $\sim 1.3$ in Hubble types $T \lesssim 1.5$.
	    \vspace{5pt}
    \item The  Type II outer discs of late-type ($T > 5$) and low mass ($\log_{10} (M_{*}/M_{\sun}) \lesssim 10.5$) galaxies show a decrease in the scalelength of the outer disc ($h_{\rm o}$) with decreasing wavelength. The steeper outer discs at shorter wavelengths, probing younger stellar populations, could be a manifestation of star formation threshold creating the break. In addition, the outer discs are populated by older stars that have migrated from the inner disc.
        \vspace{5pt}
	\item The scalelengths of the outer discs ($h_{\rm o}$) of Type III profiles are on average slightly larger in \emph{u}-band compared to infrared (a factor of $\sim 1.2$). This increase in $h_{\rm o}$ might be due to enhanced star formation in the outer discs, which could be a result of a past galaxy interaction, or gas inflow from intergalactic medium.
\end{enumerate}


\begin{acknowledgements} 
We thank the referee, Michael Pohlen, for the report that significantly improved the quality of the paper. The authors wish to thank the entire S$^4$G team for their effort. J.L. gratefully acknowledges financial support from the Vilho, Yrj\"o ja Kalle V\"ais\"al\"a foundation of the Finnish Academy of Science and Letters. J.L, E.L, and H.S acknowledge the support from Academy of Finland. We acknowledge financial support from the People Programme (Marie Curie Actions) of the European Union's FP7 2007-2013 to the DAGAL network under REA grant agreement number PITN-GA-2011-289313. 

This research is based in part on observations made with the Spitzer Space Telescope, which is operated by the Jet Propulsion Laboratory, California Institute of Technology under a contract with NASA. We are grateful to the dedicated staff at the Spitzer Science Center for their help and support in planning and execution of this Exploration Science program. We also gratefully acknowledge support from NASA JPL/Spitzer grant RSA 1374189 provided for the S$^4$G project.

This research has made use of the NASA/IPAC Extragalactic Database (NED) which is operated by the Jet Propulsion Laboratory, California Institute of Technology, under contract with the National Aeronautics and Space Administration. This research has made use of SAOImage DS9, developed by Smithsonian Astrophysical Observatory.
\end{acknowledgements}

\bibliographystyle{aa} 
\bibliography{references} 


\begin{appendix}

\section{Comparison with previous studies}

We compare our results with those previous disc break studies of low redshift galaxies which use similar methods to study the disc profiles. Our infrared sample shares 118 galaxies in common with the optical studies by \cite{pohlen2006}, \cite{erwin2008}, and \cite{gutierrez2011} (P06, E08, and G11 respectively here). We also compared with \cite{munozmateos2013}, which has 152 galaxies in common with this study, using the same 3.6 $\mu m$ images as used by us.

\begin{figure*}
    \begin{center}
        \includegraphics[width=0.9\textwidth]{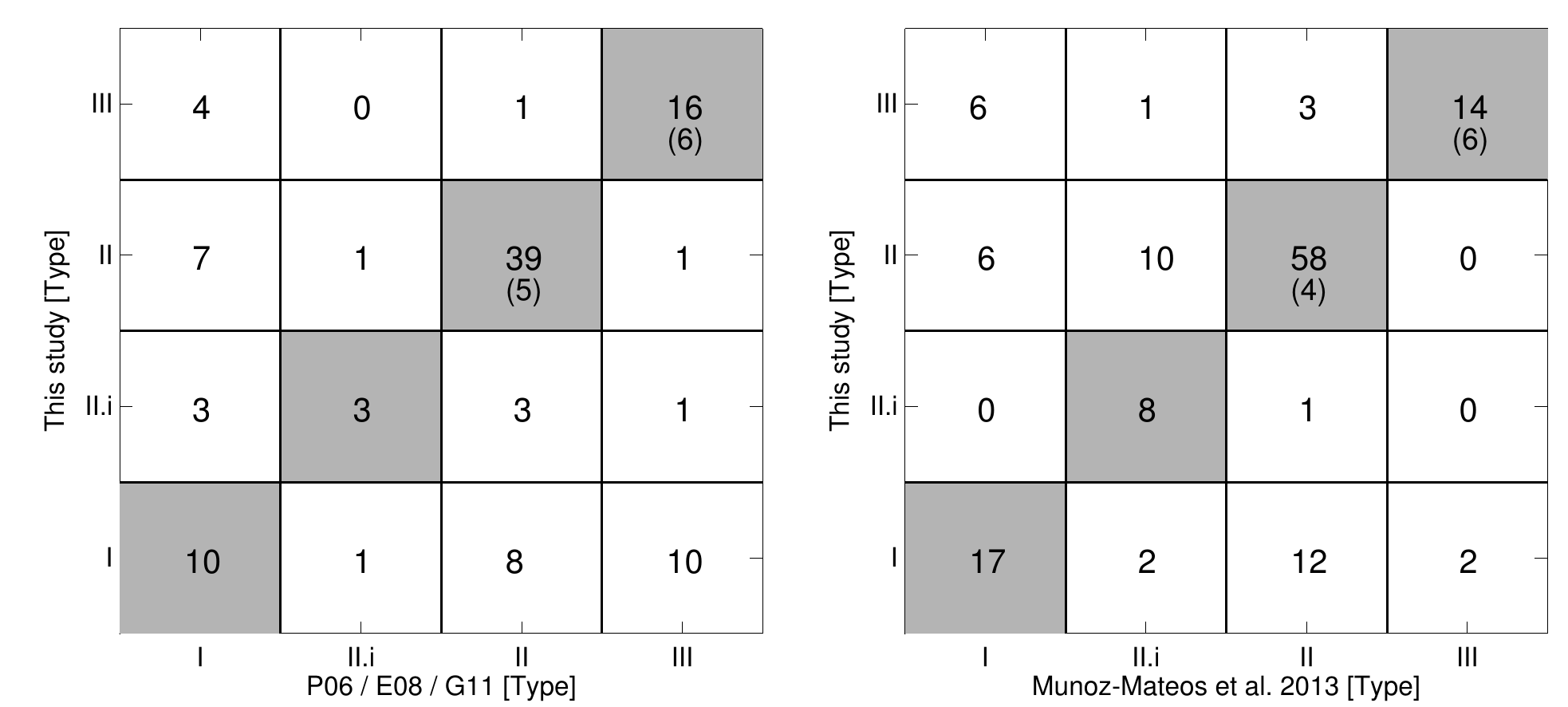}
        \caption{Break type classifications compared for galaxies that are in our study and in previous studies, namely \citeauthor{pohlen2006} (2006, P06), \citeauthor{erwin2008} (2008, E08), \citeauthor{gutierrez2011} (2011, G11), and \cite{munozmateos2013}. The values in parenthesis are additional cases in which the same type of break has been found in both studies, but either study has also found an additional break of this type in the same galaxy.}
        \label{type_diff}
        \vspace*{20pt}
        \includegraphics[width=\textwidth]{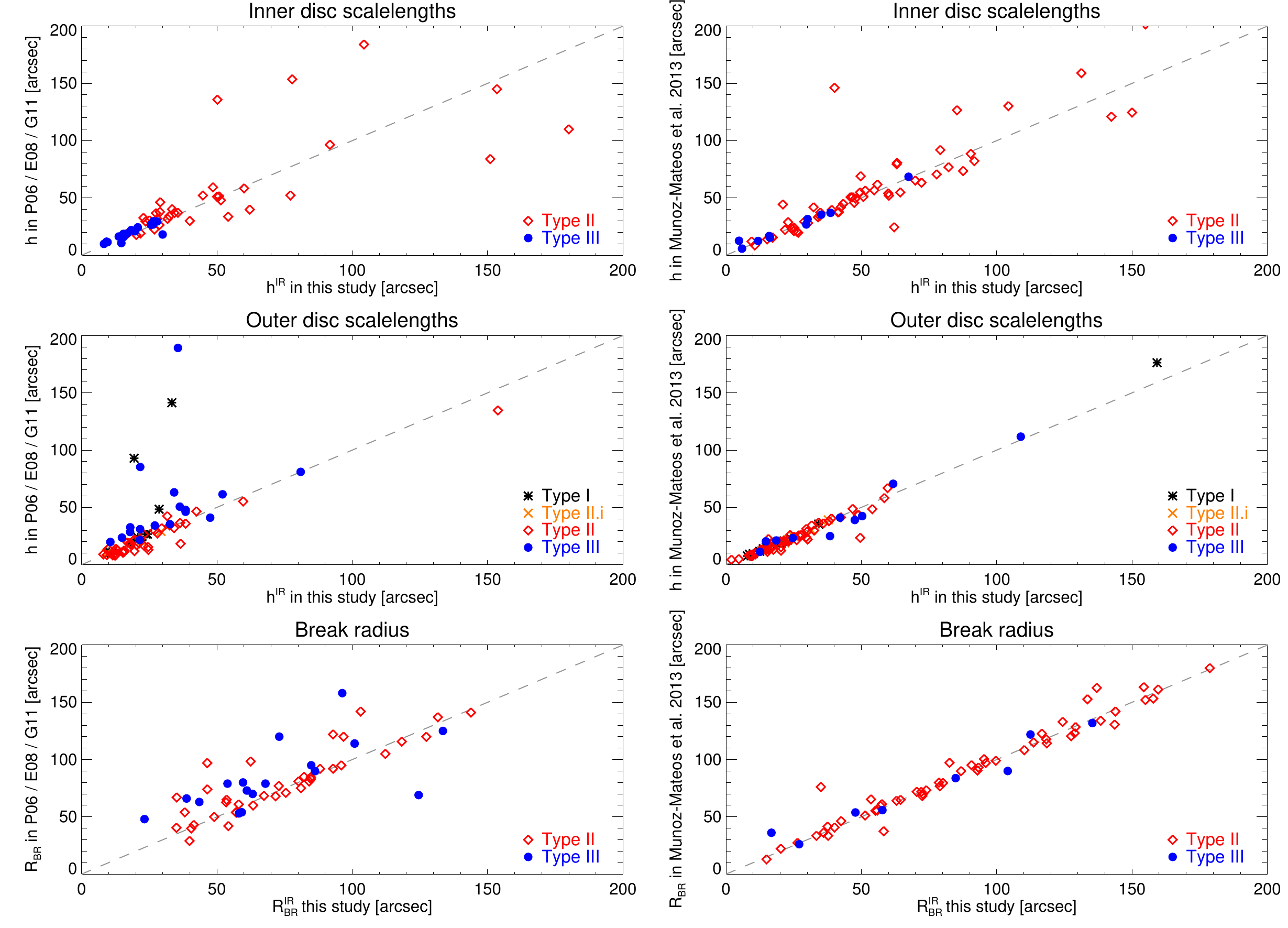}
        \caption{Disc and break parameters compared for galaxies that are in our study and in previous studies, and have also the same profile type among the different studies. We compare our parameters with \citeauthor{pohlen2006} (2006, P06), \citeauthor{erwin2008} (2008, E08), and \citeauthor{gutierrez2011} (2011, G11) in the \emph{left panels}, and with \cite{munozmateos2013} in the \emph{right panels}. }
        \label{h_trad_diff}
    \end{center}
\end{figure*}

We find that less than two thirds of the galaxies have been classified with the same profile type among the different studies. Namely, 58\% of the galaxies we share with P06, E08, and G11 (68 out of 117), and 63\% of those which are studied by \cite{munozmateos2013} (95 out of 152). In Figure \ref{type_diff} we summarize the comparison of the profile type classifications between our study and these above studies. The most common differences appear in galaxies have been classified as single exponential Type I in one study, and as Type II or III in some other study. In ten cases Type I profile in our study is classified as Type III in P06, E08, or G11 (see Fig. \ref{type_diff} left panel). In these cases the Type III break is found at particularly low surface brightness level, with average $\mu_{BR} \sim 25.2$ in the \emph{r} band, which is close to the surface brightness limit of our study (see Table \ref{table:sigmasky}). On the other hand, \cite{munozmateos2013}, using the same data as we use, classifies some galaxies with a Type II break, which we have considered to be of Type I (see Fig. \ref{type_diff} right panel). Moreover, \cite{munozmateos2013} considers nine cases as Type II.i which we have classified as normal Type II. This latter discrepancy could be explained by the different sources of bar measurements used: \cite{munozmateos2013} uses their own measurements, whereas we use the bar measurements from \cite{laurikainen2011}, \cite{herreraendoqui2015}, and \cite{diazgarcia2016}. Most probably these nine cases are galaxies in which the bar detection and size measurements have large uncertainties. The \cite{munozmateos2013} study shares 32 galaxies common with P06, E08, or G11: 17 of these galaxies have the same disc profile type ($53$\%). To conclude, the differences in the profile type classifications appear to rise mainly from two factors: subjectivity of the profile type classification because no definite criteria has been established, and differences rising from the surface brightness limit of used data. 

We then compare the parameters of the disc of the galaxies where the profile type classification is the same in our study and in P06, E08, or G11 (Fig. \ref{h_trad_diff} left panels), or in our study and in \cite{munozmateos2013} (Fig. \ref{h_trad_diff} right panels). In most cases the measurements of this study agree with those obtained by P06, E08, G11 in the optical. The agreement with the disc scalelengths and the break radii is better when compared with \cite{munozmateos2013} (Fig. \ref{h_trad_diff} right panels). This is expected because the data used is the same. In summary, if the disc profile type is the same across the studies then the disc and break parameters are also comparable. However, even when the same methods are used, the parameters returned from the disc fitting can be affected by several factors (e.g. differences in the position angle and ellipticity of the isophotes over which the surface brightness profile is constructed, and differences in the disc fitting region), which could explain the deviation of the parameters between the different studies.

\end{appendix}

\end{document}